\documentclass[aps,prc,twocolumn,showpacs,superscriptaddress]{revtex4-1}

\usepackage{subfigure}
\usepackage{graphicx, color}% Include figure files
\usepackage{float}
\usepackage{hyperref}
\usepackage[all]{hypcap}
\usepackage{array}
\usepackage{graphicx}
\usepackage{amsmath,amsthm,amssymb}
\usepackage{color}

\def\lsim{\,\raise0.3ex\hbox{$<$\kern-0.75em\raise-1.1ex\hbox{$\sim$}}\,}
\def\gsim{\,\raise0.3ex\hbox{$>$\kern-0.75em\raise-1.1ex\hbox{$\sim$}}\,}

\newcommand{\nn}{\nonumber\\}
\newcommand{\ba}{\begin{eqnarray}}
\newcommand{\ea}{\end{eqnarray}}
\newcommand{\la}[1]{\label{#1}}

\begin{document}

%\title{Conspiracy behind the $\sqrt{s}$ and $p_T$ dependence of single inclusive jet suppression in ultrarelativistic heavy-ion collisions}
\title{Interplaying mechanisms behind single inclusive jet suppression in heavy-ion collisions}

\author{Yayun He}
\affiliation{Key Laboratory of Quark and Lepton Physics (MOE) and Institute of Particle Physics, Central China Normal University, Wuhan 430079, China}
\affiliation{Nuclear Science Division Mailstop 70R0319,  Lawrence Berkeley National Laboratory, Berkeley, California 94720}
\author{Shanshan Cao}
\affiliation{Department of Physics and Astronomy, Wayne State University, Detroit, Michigan 48201}
\author{Wei Chen}
\affiliation{Key Laboratory of Quark and Lepton Physics (MOE) and Institute of Particle Physics, Central China Normal University, Wuhan 430079, China}
\author{Tan Luo}
\affiliation{Key Laboratory of Quark and Lepton Physics (MOE) and Institute of Particle Physics, Central China Normal University, Wuhan 430079, China}
\author{Long-Gang Pang}
\affiliation{Physics Department, University of California, Berkeley, California 94720}
\affiliation{Nuclear Science Division Mailstop 70R0319,  Lawrence Berkeley National Laboratory, Berkeley, California 94720}
\author{Xin-Nian Wang}
%\email{xnwang@lbl.gov}
\email[Correspondence:]{xnwang@lbl.gov}
%\homepage[]{Your web page}
%\thanks{xnwang@lbl.gov}
%\altaffiliation{}
\affiliation{Key Laboratory of Quark and Lepton Physics (MOE) and Institute of Particle Physics, Central China Normal University, Wuhan 430079, China}
\affiliation{Nuclear Science Division Mailstop 70R0319,  Lawrence Berkeley National Laboratory, Berkeley, California 94720}
\affiliation{Physics Department, University of California, Berkeley, California 94720}

\begin{abstract}
The suppression factor for single inclusive jets in Pb+Pb collisions at the Large Hadron Collider (LHC) has a weak dependence on the transverse momentum $p_T$ and remains almost the same at two colliding energies,  $\sqrt{s}=2.76$ and 5.02 TeV, though the central rapidity density of bulk hadrons increases by about 20\%.  This phenomenon is investigated within the Linear Boltzmann Transport (LBT) model, which includes elastic and inelastic processes based on perturbative QCD for both jet shower and recoil medium partons as they propagate through a quark-gluon plasma (QGP). With the dynamic evolution of the QGP given by the 3+1D  CLVisc hydrodynamic model with event-by-event fully fluctuating initial conditions, single inclusive jet suppression in Pb+Pb collisions from LBT agrees well with experimental data.  The weak $\sqrt{s}$ and  $p_T$-dependence of the jet suppression factor at LHC are found to result directly from the $\sqrt{s}$-dependence of the initial jet $p_T$ spectra and slow $p_T$-dependence of the jet energy loss. Contributions from jet-induced medium response, influence of radial expansion, both of which depend on jet-cone size, and jet flavor composition all conjoin to give a slow $p_T$-dependence of jet energy loss and the single jet suppression factor $R_{\rm AA}$, their dependence on $\sqrt{s}$ and jet-cone size. Single inclusive jet suppression at $\sqrt{s}=200$ GeV is also predicted that actually decreases slightly with $p_T$ in the $p_T<50$ GeV/$c$ range because of the steeper initial jet spectra though the $p_T$-dependence of the jet energy loss is weaker than that at LHC.
\end{abstract}

\maketitle

\section{Introduction}
\label{intro}

Jet quenching caused by parton energy loss in dense medium has been proposed as a hard probe of the properties of the quark-gluon plasma (QGP) formed in high-energy heavy-ion collisions \cite{Bjorken:1982tu,Gyulassy:1990ye}. The simplest form of jet quenching is the suppression of single inclusive hadron spectra at large transverse momentum, dihadron and $\gamma$-hadron correlation in heavy-ion collisions relative to proton-proton collisions \cite{Wang:1991xy,Wang:1996yh,Vitev:2002pf,Wang:2003mm,Eskola:2004cr,Majumder:2004pt,Renk:2006nd,Qin:2007rn,Zhang:2007ja,Zhang:2009rn,Majumder:2010qh,Zapp:2012ak,Qin:2015srf}.  Observation of these jet quenching phenomena among other experimental data on collective phenomena at the Relativistic Heavy-Ion Collider (RHIC) provided the first evidence of the formation of the strongly coupled quark-gluon plasma in high-energy heavy-ion collisions \cite{Adcox:2004mh,Adams:2005dq}. A systematic study of experimental data on suppression of single inclusive hadron spectra in heavy-ion collisions at both RHIC and the Large Hadron Collider (LHC) has provided unprecedented constraints on jet transport coefficients \cite{Burke:2013yra}.

Since the inclusive hadron spectra at large transverse momentum $p_T$ is the convolution of cross sections of energetic parton production and parton fragmentation functions in which leading hadrons dominate, the suppression of single inclusive hadron spectra is caused mainly by the energy loss of leading jet partons inside the dense QGP medium that suppresses the effective jet fragmentation functions at large momentum fraction. The hadron suppression factor is therefore not sensitive to the distribution of soft radiative gluons and recoil partons from jet-induced medium response. This is, however, not the case for fully reconstructed jets in heavy-ion collisions.

Suppression and modification of full jets are also proposed to study jet quenching and properties of QGP medium in high-energy heavy-ion collisions
 \cite{Vitev:2008rz,Vitev:2009rd,Qin:2010mn,He:2011pd,Neufeld:2012df,Renk:2012cx,Dai:2012am,Qin:2012gp,Wang:2013cia,Casalderrey-Solana:2014bpa,Chang:2016gjp,Casalderrey-Solana:2016jvj,Milhano:2015mng,Kang:2017frl}.  
  Jets are collimated clusters of hadrons within a given cone-size in experimental measurements. In elementary hadronic processes such as proton-proton collisions, the jet production cross section can be calculated from perturbative QCD (pQCD) and can describe experimental data to high precision even with relatively small cone sizes \cite{Sterman:1977wj,Cacciari:2008gp,Kang:2017frl}. The cross section is not very sensitive to nonperturbative processes of jet hadronization through fragmentation. In heavy-ion collisions, however, the final jet production cross section is not only modified by parton energy loss of leading partons but also is influenced by how the lost energy is transported in the medium through radiated gluons and recoil medium partons. It is therefore imperative to include the effect of recoil partons and their further propagation in the form of jet-induced medium response as well as the propagation of radiated gluons in the study of jet suppression and medium modification \cite{Wang:2013cia,Tachibana:2015qxa,Casalderrey-Solana:2016jvj,Wang:2016fds,Tachibana:2017syd,KunnawalkamElayavalli:2017hxo,Milhano:2017nzm,Chen:2017zte,Luo:2018pto}.
 
Contributions from jet-induced medium response to the jet energy within a finite jet-cone size should also be influenced by the collective radial expansion and flow of the medium. They will affect the transverse momentum $p_T$ dependence of jet energy loss in heavy-ion collisions. Since the interaction strengths of gluon and quark with the medium are different due to their color charges, one should also expect a flavor dependence of the jet energy loss. The fractions of gluon and quark jets and their $p_T$ and colliding energy $\sqrt{s}$ dependence are determined by the pQCD cross sections and initial parton distributions in the colliding nuclei. All these conjoin to give a particular $\sqrt{s}$ and $p_T$ dependence of the jet energy loss that can explain the observed phenomenon in the suppression of single inclusive jets in heavy-ion collisions at LHC. The suppression factor for single inclusive jets has been measured in Pb+Pb collisions at two colliding energies, $\sqrt{s}=2.76$ and 5.02 TeV, at LHC~\cite{Aad:2014bxa,Aaboud:2018twu,Khachatryan:2016jfl}. The measured suppression factor has a weak $p_T$ dependence and remains almost the same at two colliding energies though the central rapidity density of bulk hadrons increases by about 20\%~\cite{Abbas:2013bpa,Adam:2016ddh}. 

In this paper, we will use the Linear Boltzmann Transport (LBT) model \cite{Li:2010ts,Wang:2013cia,He:2015pra,Cao:2016gvr,Cao:2017hhk,Luo:2018pto} for jet interaction and propagation in dense QGP medium to study the suppression of single inclusive jet spectra in high-energy heavy-ion collisions. We will pay particular attention to effects of recoil thermal partons and their further propagation in the dense medium whose evolution in high-energy heavy-ion collisions is described by a 3+1D viscous relativistic hydrodynamic model. We will try to understand the weak transverse momentum and colliding energy dependence of the suppression factor for single inclusive jet spectra in Pb+Pb collisions at LHC energies. We will investigate the effect of recoil thermal partons from jet-induced medium response, their transport in the medium and influence of radial expansion on the effective jet energy loss as well as the transverse momentum and colliding energy dependence of the flavor composition of jets.  We will also provide predictions of the cone-size dependence of the jet suppression factor in Pb+Pb collisions at LHC and jet suppression in Au+Au collisions at RHIC energy $\sqrt{s}=200$ GeV.

The remainder of this paper is organized as follows. We will provide a brief description of the LBT model and simulations of jet propagation in the dense medium whose evolution in high-energy heavy-ion collisions is given by the 3+1D CLVisc hydrodynamic model \cite{Pang:2012he,Pang:2014ipa,Pang:2018zzo} in Sec.~\ref{lbt}.  In Secs.~\ref{ppjet} and \ref{aajet}, we carry out calculations of the single inclusive jet spectra in both p+p collisions as the baseline and Pb+Pb collisions.  Effects of recoil medium partons, diffusion wake due to the back-reaction and underlying event subtractions are studied in detail.  Results on single inclusive jet suppression in Pb+Pb collisions at other centralities and at both $\sqrt{s}$=2.76 and 5.02 TeV are presented and compared to experimental data.  Section~\ref{sec:jetsuppression} is devoted to the discussion and understanding of the colliding energy and jet transverse momentum dependence of the jet suppression in heavy-ion collisions. In Sec.~\ref{sec:eloss}, we examine in detail effects of transport of recoil partons, radial expansion of the underlying bulk medium and the flavor composition on the effective jet energy loss in heavy-ion collisions. These effects combined with the shape and colliding energy dependence of the initial jet production spectra in p+p collisions can explain the weak transverse momentum and colliding energy dependence of the single inclusive jet suppression factor. They also lead to a unique cone-size dependence of the jet suppression. We will also provide predictions for single inclusive jet suppression at the RHIC energy $\sqrt{s}=200$ GeV in Sec.~\ref{rhicpredict}. A summary and discussion are given in Sec.~\ref{summary}.

\section{The Linear Boltzmann Transport model}

\label{lbt}

The Linear Boltzmann Transport (LBT) model is developed to study jet interaction and propagation in dense QGP medium with a particular emphasis on
thermal recoil partons and their further interaction and propagation through the medium in the form of jet-induced medium excitation (or response). It was
initially developed \cite{Li:2010ts} to study the so-called Mach-cone excitation by jets that travel at nearly the speed of light in the medium in which the velocity of sound is smaller than that of the propagating jets \cite{CasalderreySolana:2004qm,Stoecker:2004qu,Ruppert:2005uz,Chaudhuri:2005vc}.  While signals of the Mach-cone excitation are still elusive in both experimental measurements and simulations with realistic hydrodynamic evolution of the medium, the LBT model becomes a powerful tool for the study of jet quenching in high-energy heavy-ion collisions. The model has been recently improved with the implementation of the complete set of elastic $2\to 2$ scattering processes \cite{He:2015pra}. Inelastic processes  $2\rightarrow 2+n$ with multiple gluon radiation and global energy-momentum conservation have also been implemented more consistently in the latest version \cite{Wang:2013cia,Luo:2016ufs,Cao:2016gvr}. It has been used to describe both single inclusive light and heavy flavor hadron suppression \cite{Cao:2017hhk}, $\gamma$-hadron \cite{Chen:2017zte}, $\gamma$-jet \cite{Wang:2013cia,Luo:2016ruj,Luo:2018pto} and $Z^0$-jet correlations~\cite{Zhang:2018urd}. We will use it to study single inclusive jet suppression in high-energy heavy-ion collisions in this paper.

The basic building block of the LBT model is the linear Boltzmann equations for the transport of both jet shower and thermal recoil partons in QGP,
\begin{eqnarray}
p_a\cdot\partial f_a&=&\int \sum_{b c d } \prod_{i=b,c,d}\frac{d^3p_i}{2E_i(2\pi)^3} (f_cf_d-f_af_b)|{\cal M}_{ab\rightarrow cd}|^2
\nn && \hspace{-0.5in}\times \frac{\gamma_b}{2}
S_2(\hat s,\hat t,\hat u)(2\pi)^4\delta^4(p_a\!+\!p_b\!-\!p_c\!-\!p_d)+ {\rm inelastic},
\label{bteq}
\end{eqnarray}
where the summation is over all possible parton flavors and channels of scattering, $f_i=(2\pi)^3\delta^3(\vec{p}-\vec{p_i})\delta^3(\vec{x}-\vec{x_i}-\vec{v_i}t)$ $(i=a,c)$ are the phase-space density for jet shower partons before, after scattering and medium recoil partons,  $f_i=1/(e^{p_i\cdot u/T}\pm1)$ $(i=b,d)$ are phase-space distributions for thermal partons in the QGP medium with local temperature $T$ and fluid velocity $u=(1, \vec{v})/\sqrt{1-\vec{v}^2}$, and $\gamma_b$ is the color-spin degeneracy for parton $b$.

The leading-order (LO) elastic scattering amplitudes $|{\cal M}_{ab\rightarrow cd}|^2$ \cite{Eichten:1984eu} have collinear divergencies that are
regularized in the LBT model by a factor \cite{Auvinen:2009qm},
\begin{equation}
S_2(\hat s, \hat t, \hat u) = \theta(\hat s\ge 2\mu_{D}^2)\theta(-\hat s+\mu_{D}^2\le \hat t\le -\mu_{D}^2),
\end{equation}
 where $\hat s$, $\hat t$, and $\hat u$ are Mandelstam variables, and 
 \begin{equation}
 \mu_{D}^2 = \frac{3}{2}g^2 T^2,
 \label{eq-mud}
 \end{equation}
 is the Debye screening mass with three quark flavors. The corresponding elastic cross sections are $d\sigma_{ab\rightarrow cd}/d\hat t=|{\cal M}_{ab\rightarrow cd}|^2/16\pi \hat s^2$. We neglect the Bose enhancement (Pauli blocking) for final-state gluons (quarks) and detailed balance of the radiative processes in the current implementation of the Boltzmann transport. The strong coupling constant $\alpha_s=g^{2}/4\pi$ is fixed and will be fitted to experimental data.

In the current version of the LBT model,  we only consider gluon radiation induced by elastic scatterings. The differential inclusive rates for gluon radiation is assumed to follow that from the high-twist approach \cite{Guo:2000nz,Wang:2001ifa},
\ba \la{induced}
\frac{d\Gamma_{a}^{\rm inel}}{dzdk_\perp^2}=\frac{6\alpha_sP_a(z)k_\perp^4}{\pi (k_\perp^2+z^2m^2)^4} \frac{p\cdot u}{p_0}\hat{q}_{a} (x)\sin^2\frac{\tau-\tau_i}{2\tau_f},
\ea
where $P_a(z)$ is the splitting function for the propagating parton $a$ to emit a gluon with the energy fraction $z$ and transverse momentum $k_\perp$, $m$ is the mass of the propagating parton, $\tau_f=2p_0z(1-z)/(k_\perp^2+z^2m^2)$ is the gluon formation time and $\tau_i$ is the time of the last gluon emission.
The elastic scattering rate in the inelastic processes has been factorized into the jet transport coefficient,
\begin{equation}
\hat{q}_{a}(x)=\sum_{bcd}\rho_{b}(x)\int d\hat t q_\perp^2 \frac{d\sigma_{ab\rightarrow cd}}{d\hat t},
\label{eq-qhat}
\end{equation}
which is defined as the transverse momentum transfer squared per mean-free-path in the local comoving frame of the QGP fluid.  The parton density $\rho_{b}(x)$ includes the degeneracy factor. The splitting functions $P_a(z)$ above contains an infrared divergence and is regularized by the Debye screening mass $\mu_D$ as an infrared cut-off for the energy of radiated gluons.

In the actual implementation of parton transport simulations in LBT, the probability of elastic and inelastic scattering in each small but finite time step $\Delta \tau $ are calculated together to ensure unitarity. The probability for an elastic scattering in a time step  $\Delta \tau $ during the propagation of parton $a$ is,
\begin{equation}
P^a_{\rm el}=1-\text{exp}[- \Delta\tau \Gamma_a^{\rm el}(x)],
\end{equation}
where
\begin{equation}
\Gamma_a^{\rm el}\equiv \frac{p\cdot u}{p_0}\sum_{bcd} \rho_b(x)\sigma_{ab\rightarrow cd}
\label{eq-rate}
\end{equation}
is the total elastic scattering rate for parton $a$.  The probability for inelastic process is
\begin{equation}
P^a_\mathrm{inel}=1-\exp[-\Delta\tau \Gamma_a^{\rm inel}(x)], 
\end{equation}
where
\begin{equation}
\Gamma_a^{\rm inel}=\frac{1}{1+\delta_g^a}\int dz dk_\perp^2 \frac{d\Gamma_a^{\rm inel}}{dzdk_\perp^2}
\end{equation}
is the total gluon radiation rate from parton $a$. The total scattering probability,
\begin{equation}
P^a_\mathrm{tot}=P^a_\mathrm{el}(1-P^a_\mathrm{inel}) +P^a_\mathrm{inel},
\end{equation}
can be separated  into the probability for pure elastic scattering (first term) and that for inelastic scattering with at least one gluon radiation (the second term).  Notice that for infinitesimally small time step $\Delta\tau \rightarrow 0$, the above total scattering probability per unit time is just the sum of the elastic and inelastic scattering rate. 

A Poisson distribution with the mean $\langle N^a_g \rangle=\Delta\tau\Gamma_a^{\rm inel}$ is assumed to simulate multiple gluon radiations associated with each elastic scattering. The scattering channel, flavor, energy and momentum of the final partons, recoil partons and radiated gluons are sampled according to the differential elastic scattering cross section and the differential gluon radiation rate, respectively. Global energy and momentum conservation is ensured in each scattering with multiple radiated gluons.

In the LBT model, the above scattering probabilities are employed to simulate the change of phase-space distribution for jet shower, recoil medium partons and radiated gluons due to their scattering with thermal partons in the medium. During each scattering, the initial thermal parton $b$ is recorded as ``negative'' partons and they are also allowed to propagate in the medium according to the Boltzmann equation. The energy and momentum of these ``negative" partons will be subtracted from all final observables to account for the back-reaction in the Boltzmann transport equations. They are part of the jet-induced medium excitation and manifest as the diffusion wake behind the propagating jet shower partons \cite{Wang:2013cia,Li:2010ts,He:2015pra}.

In the LBT model we assume jet shower parton density and jet-induced medium response is small in the linear approximation ($\delta f\ll f$) so that one can neglect interaction among jet shower and recoil partons. One considers only interaction between jet shower and recoil partons with thermal medium partons. The bulk medium evolves independently according to a hydrodynamic model that provides spatial and time information on the local temperature and fluid velocity during parton-medium interaction. This linear approximation will break down when the jet-induced medium excitation becomes comparable to the local thermal parton density.  To extend LBT beyond the linear approximation, a coupled LBT and hydrodynamic (CoLBT-hydro) model \cite{Chen:2017zte} has been developed in which soft partons from LBT jet transport are fed back to the bulk medium as a source term in the hydrodynamic equations while energetic partons propagate through the medium which evolve simultaneously with the source term updated in real time. This coupled approach is important for detailed study of the jet-induced medium excitation. For the study of jet suppression, the LBT model with the linear approximation will suffice.

In the LBT model, a parton recombination model developed by the Texas A \& M University group within the JET Collaboration~\cite{Han:2016uhh} is used for hadronization of both jet shower and recoil medium partons. The model has been used successfully to describe light flavor hadron suppression in heavy-ion collisions \cite{Chen:2017zte}. In this paper, we will only use the partonic information for jet reconstruction and study single inclusive jet suppression and jet energy loss.

\section{Single inclusive jet spectra in p+p collisions}
\label{ppjet}

For the study of single inclusive jet spectra in high-energy heavy-ion collisions, we have to first provide initial jet shower parton distributions from elementary nucleon-nucleon collisions and then let these jet shower partons propagate in the LBT model through bulk medium that evolves according to the hydrodynamic model. Each of the initial jet shower partons is assigned with a formation time determined from their virtuality, energy and transverse momentum (relative to the jet direction). They start interaction with medium partons only after their initial formation time.   We then use the information for the final partons  and the FASTJET package \cite{Cacciari:2011ma}, which is specially modified to take into account of the subtraction of  ``negative" partons, with the anti-$k_{t}$ algorithm to reconstruct jets and calculate the final single inclusive jet spectra. 

We will use PYTHIA 8 \cite{Sjostrand:2006za} to simulate production of initial jet shower partons in this study. To ensure enough statistics for initial jet production at any large transverse momentum, we divide the range of transverse momentum to many bins with bin size $dp_{T i}$. We then use PYTHIA 8 to generate initial jet shower partons (with both initial and final state radiation) with a trigger on the transverse momentum transfer $p_{Ti} \in (p_{T i}-d p_{T i}/2, p_{T i}+dp_{T i}/2)$ and the cross section $d\sigma_{\rm LO}^{{\rm pp}(c)}/dp_{T i}$ in the leading-order (LO) perturbative QCD (pQCD) for production of initial hard parton $c$ in p+p collisions. For any given trigger $p_{T i}$, we generate a given number of events for jet production. After jet reconstruction using FASTJET with a given jet-cone radius $R$, one can get an event-averaged single inclusive jet distribution $dN^{\rm jet}_{(c)}(p_{Ti})/dydp_T$ for a given trigger $p_{Ti}$, here $p_T$ and $y$ are the transverse momentum and rapidity of the final jet, respectively,  as reconstructed from the final partons with FASTJET. The final single inclusive jet cross section in p+p collisions will be given by,
\begin{equation}
    \frac{d^2\sigma^{\rm jet}_{\rm pp}}{dp_Tdy} = \sum_c\int dp_{T i}  \frac{d\sigma_{\rm LO}^{{\rm pp}(c)} }{dp_{Ti}}
     \frac{d^2N^{\rm jet}_{(c)}(p_{Ti}, p_T)} {dp_T dy},
    \label{eq-jetcrs}
\end{equation}
where the LO pQCD cross section for the production of initial hard parton $c$ in p+p collisions is given by
\begin{eqnarray}
\frac{d \sigma^{{\rm pp}(c)}_{\rm LO}}{dp_{Ti}} & = & 2 p_{Ti}\sum_{a,b,d}  \int dy_c dy_d  x_a f_{a/p} (x_a, \mu^2) 
 \nonumber\\ 
 & &  \times x_b f_{b/p} (x_b, \mu^2) \frac{d\hat\sigma_{ab\to cd}}{dt},
\label{eq:cs.pp}
\end{eqnarray}
where $y_c$ and $y_d$ are rapidities of the final hard partons in the $a+b\rightarrow c+d$ processes, $x_a=x_{Ti}(e^{y_c}+e^{y_d})$ and $x_b=x_{Ti}(e^{-y_c}+e^{-y_d})$ are the light-cone momentum fractions carried by the initial partons from the two colliding protons with $x_{Ti}=2p_{Ti}/\sqrt{s}$, $f_{a/p}(x,\mu^2)$ is the parton distribution function inside a proton at the scale $\mu^2=p_{Ti}^2$ and 
$d\hat\sigma_{ab\to cd}/dt$ is the parton level leading order cross section which depends on the Mandelstam variables 
$\hat s=x_ax_bs$, $\hat t=-p_{Ti}^2(1+e^{y_d-y_c})$ and $\hat u=-p_{Ti}^2(1+e^{y_c-y_d})$.  Because of higher-order corrections through initial and final state radiation in PYTHIA 8, there can be more than two jets in the final state and the transverse momentum $p_T$ of the final leading jets can be different from the value of the trigger $p_{Ti}$.  

\begin{figure}[htbp]
    \centering
    \includegraphics[width=8.5cm]{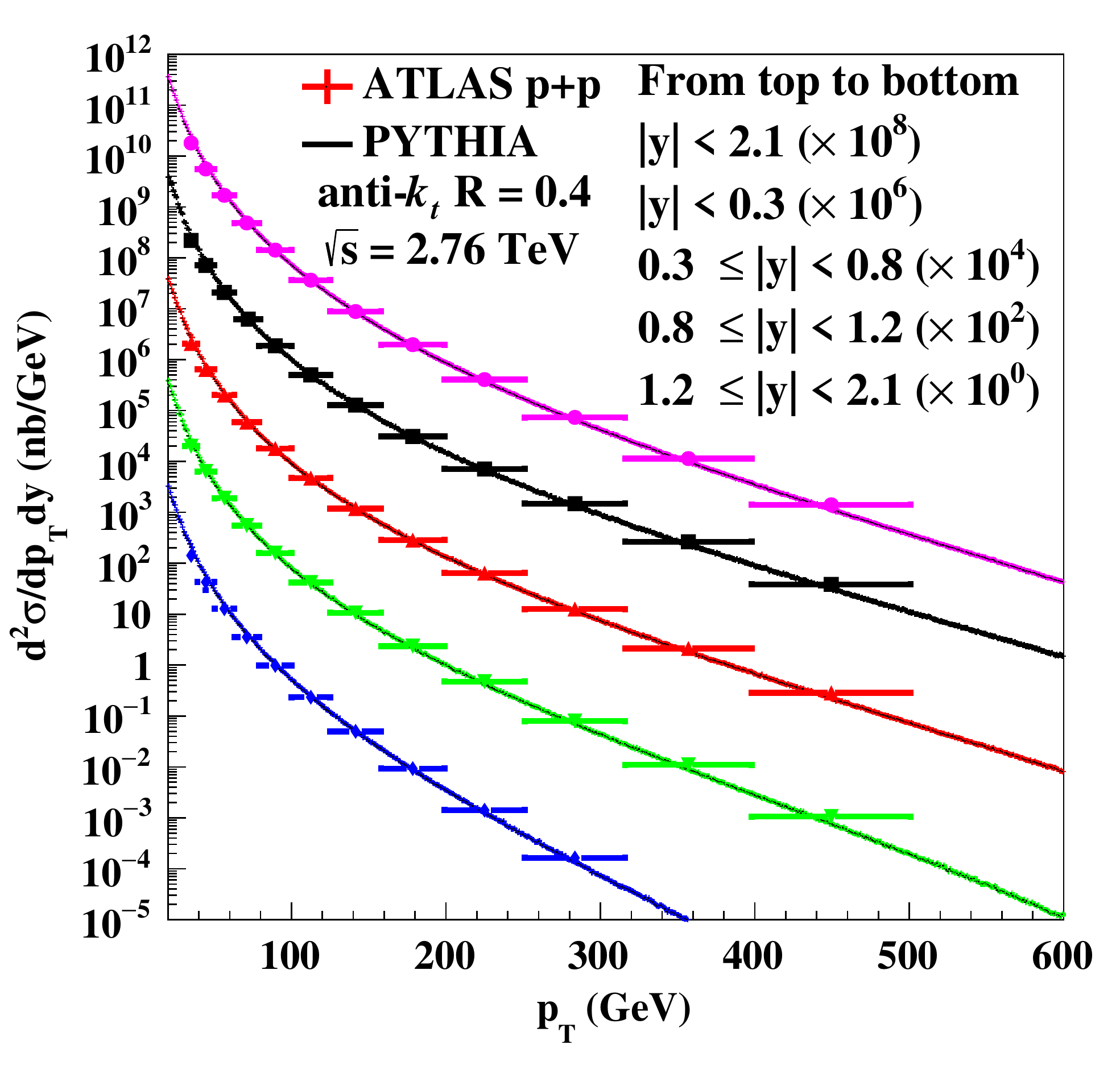}
    \caption{(Color online) The single inclusive jet double differential cross section as a function of $p_{T}$ in different rapidity bins in p+p collisions at $\sqrt{s} = 2.76$ TeV using anti-$k_{t}$ algorithm with jet cone radius R = $0.4$. The closed symbols are ATLAS experimental data \cite{Aad:2014bxa} while the curves are from PYTHIA 8 simulations. The results for different rapidities are scaled by successive powers of $10^{2}$ for clear presentation.}
    \label{jetCS_2760}
\end{figure}

\begin{figure}[htbp]
    \centering
    \includegraphics[width=8.5cm]{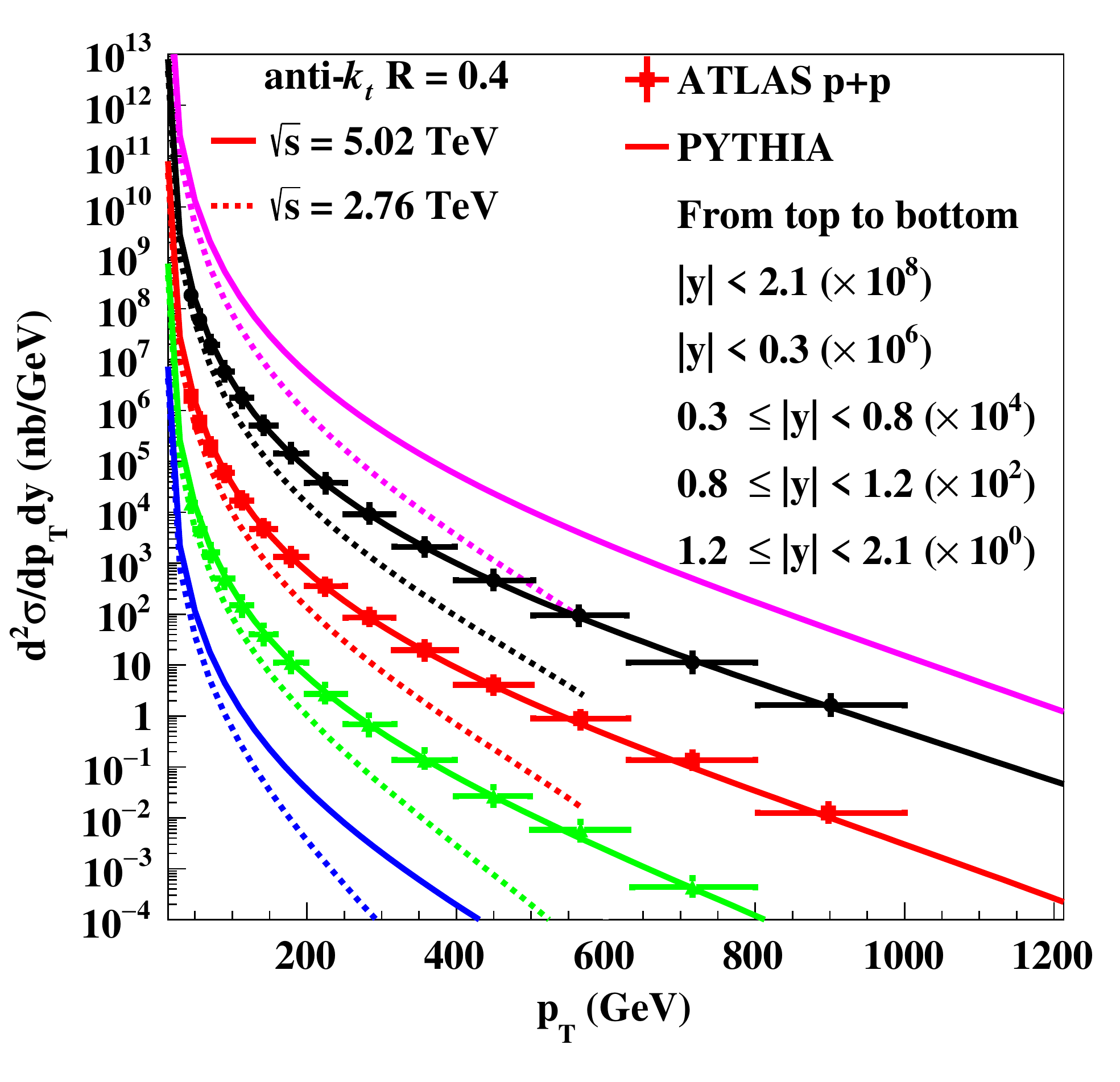}
    \caption{(Color online) The inclusive jet double differential cross section as a function of $p_{T}$ in different rapidity bins in p+p collisons at $\sqrt{s} = 5.02$ TeV (solid) using anti-$k_{t}$ algorithm with jet cone radius R = $0.4$ from PYTHIA 8 as compared to ATLAS experimental data \cite{Aaboud:2018twu}. PYTHIA 8 results at $\sqrt{s} = 2.76$ (dashed) are also shown as a comparison. Results for different rapidities are scaled by successive powers of $10^{2}$.}
    \label{jetCS_twoEnergy}
\end{figure}

Shown in Figs.~\ref{jetCS_2760} and \ref{jetCS_twoEnergy} are differential single inclusive jet cross sections with jet-cone size $R=0.4$ as a function of the final jet transverse momentum $p_T$ in different rapidity windows of p+p collisions at $\sqrt{s}=2.76$ and 5.02 TeV, respectively,  from PYTHIA 8 as compared to ATLAS experimental data \cite{Aad:2014bxa,Aaboud:2018twu}.  PYTHIA 8 can describe the experimental data well. In Fig.~\ref{jetCS_twoEnergy}, we also compare the single inclusive jet spectra at two different colliding energies at LHC. One can see that the shape of the single inclusive jet spectra at $\sqrt{s}=5.02$ TeV are much flatter than at 2.76 TeV, which is determined mainly by the parton distribution functions in a proton.

\section{Suppression of single inclusive jet spectra in A+A collisions}
\label{aajet}

\subsection{ Single inclusive jet cross section in A+A collisions}

We assume that the initial production rates of hard partons in A+A collisions are the same as the superposition of nucleon-nucleon collisions,
except that we need to consider the nuclear modification of the initial parton distributions  \cite{Eskola:2009uj,Ru:2016wfx}. The jet shower
partons from PYTHIA 8 simulations in each event will then go through  medium transport and propagation within the LBT model. Using FASTJET with the
same jet cone-size $R$ for jet reconstruction, we get an event-averaged final single inclusive jet distribution $d\widetilde{N}^{\rm jet}_{(c)}(p_{Ti},{\bf r},{\bf b},\phi_c)/dydp_T$ for any given transverse coordinate $\bf r$ of the binary nucleon-nucleon collision that produces the initial hard partons,  the impact-parameter $\bf b$ of the nucleus-nucleus collisions and the azimuthal angle $\phi_c$ of the initial hard parton $c$. The cross section for single inclusive jet production in A+A collision is then given by,
\begin{eqnarray}
\frac{d \sigma^{\rm jet}_{\rm AA}}{dp_{T}dy} & = &\sum_{a,b,c,d}  \int d^2{\bf r} d^2{\bf b}  t_A(r) t_A(|{\bf b}-{\bf r}|) \frac{d\phi_c}{\pi}  dy_c dy_d \nonumber\\
&& \times  \int dp_{Ti} p_{Ti}  x_a f_{a/A} (x_a, \mu^2) x_b f_{b/B} (x_b, \mu^2)
\nonumber \\
&& \times \frac{d\hat\sigma_{ab\to cd}}{dt} \frac{d\widetilde{N}^{\rm jet}_{(c)}(p_{Ti},p_T,{\bf r},{\bf b},\phi_c)}{dydp_T},
\label{eq:cs.aa}
\end{eqnarray}
where $t_{A}(r)$ is the nuclear thickness function with normalization $\int d^2{\bf r} t_A(r)=A$ and $f_{a/A}(x,\mu^2)$ is the nuclear modified parton distribution function \cite{Eskola:2009uj,Ru:2016wfx} per nucleon. The range of integration over the impact parameter $\bf b$ is determined by the centrality of the nucleus-nucleus collisions according to the experimental measurement.

Interaction between shower and medium partons in heavy-ion collisions will in general reduce the transverse momentum of the final jets, leading to the medium modification of the final single inclusive jet distribution $d\widetilde{N}^{\rm jet}_{(c)}(p_{Ti},{\bf r},{\bf b},\phi_c)/dydp_T$ relative to the vacuum one, $dN^{\rm jet}_{(c)}(p_{Ti})/dydp_T$, in p+p collisions. This will lead to the suppression of the single inclusive jet cross section in heavy-ion collisions. The suppression factor is given by the ratio of the jet cross sections for A+A and p+p collisions normalized by the averaged number of binary nucleon-nucleon collisions,
\begin{equation}
R_{\rm AA}=\frac{1}{\int d^2rd^2b  t_A(r) t_A(|{\bf b}-{\bf r}|)} \frac{d\sigma^{\rm jet}_{\rm AA}}{d\sigma^{\rm jet}_{\rm pp}}.
\label{eq:raa}
\end{equation}

In the jet reconstruction with FASTJET we also subtract underlying-event (UE) background in a scheme inspired by the method in the experimental studies \cite{Aad:2012vca}. Seed jets are defined as those with at least one particle whose transverse energy is larger than 3 GeV and with a leading particle whose transverse energy is four times or larger than the average transverse energy  per particle within the jet.  The UE background transverse energy density is calculated over the whole area of coverage excluding the area of these seed jets. In heavy-ion collisions, we also include modulation of the UE transverse energy distribution due to anisotropic flow of the bulk medium. This UE transverse energy within the transverse area of each jet is then subtracted from the jet energy in both p+p and A+A collisions. In LBT simulations, only jet shower partons, radiated gluons and recoil medium partons (energy carried by the ``negative" partons is subtracted) are used for jet reconstruction in FASTJET. The UE background is very small as compared to the UE in experimental analyses which includes all hadrons from the bulk medium. The contribution of UE to the jet energy before the subtraction in LBT simulations is about a few percent in central Pb+Pb and much smaller in p+p collisions.
The effect of UE is more important for low energy jets with large jet radii.

For heavy-ion collisions, we will use PYTHIA 8 to simulate the production of initial jet shower partons which will then propagate through the dynamically evolving  QGP medium according to the LBT model. We will neglect the nuclear modification of the initial parton distributions in cold nuclei which should be small in the jet production processes with momentum scale $Q^2> 4000$ GeV$^2$ \cite{Eskola:2009uj,Ru:2016wfx}.  We assign a formation time $\tau_0\approx 2k_0/k_T^2$ for each of the initially produced jet shower partons before which the parton is assumed to free-stream without interaction with medium partons. 

\subsection{CLVisc hydrodynamics for bulk medium evolution}

For the space-time evolution of the QGP medium in heavy-ion collisions, we use the space-time profile from the CLVisc (3+1)D viscous hydrodynamic model \cite{Pang:2014ipa,Pang:2012he}. CLVisc parallelizes the Kurganov-Tadmor algorithm \cite{KURGANOV2000241} to solve the hydrodynamic equation for the bulk medium and Cooper-Frye particlization on GPU, using Open Computing Language (OpenCL). With massive amount of computing parallelized on GPUs and Single Instruction Multiple Data (SIMD) vector operations on modern CPUs, CLVisc brings about the best performance increase so far to (3+1)D hydrodynamics on heterogeneous computing devices and provide the event-by-event space-time hydrodynamic profiles for simulations of jet transport within the LBT model in this study. The initial condition for energy-momentum density distributions for event-by-event CLVisc hydro simulations are obtained from partons in A Multi-Phase Transport (AMPT) model \cite{Lin:2004en} with a Gaussian smearing,
\begin{equation}
  \begin{aligned}
  T^{\mu\nu} &(\tau_{0},x,y,\eta_{s}) = K\sum_{i}
  \frac{p^{\mu}_{i}p^{\nu}_{i}}{p^{\tau}_{i}}\frac{1}{\tau_{0}\sqrt{2\pi\sigma_{\eta_{s}}^{2}}}\frac{1}{2\pi\sigma_{r}^{2}}\\
     	  &\hspace{-0.1in} \times \exp \left[-\frac{(x-x_{i})^{2}+(y-y_{i})^{2}}{2\sigma_{r}^{2}} - \frac{(\eta_{s}-\eta_{i s})^{2}}{2\sigma_{\eta_{s}}^{2}}\right],
  \end{aligned}
  \label{eq:Pmu}
\end{equation}
where $p^{\tau}_{i}=m_{iT}\cosh(Y_{i}-\eta_{i s})$, $p^{x}_{i}=p_{i x}$, $p^{y}_{i}=p_{i y}$ 
and $p^{\eta}_{i}=m_{i T}\sinh(Y_{i}-\eta_{i s})/\tau_{0}$ for parton $i$, which runs over all partons produced in the AMPT model simulations. 
We have chosen $\sigma_{r}=0.6$ fm, $\sigma_{\eta_{s}}=0.6$ in our calculations.
The transverse mass $m_{T}$, rapidity $Y$ and  spatial rapidity $\eta_{s}$ are calculated from the parton's four-momenta and spatial coordinates. There is no 
Bjorken scaling in the above initial condition because of early parton cascade before the initial time and the uncertainty principle applied to the initial formation time in AMPT. The scale factor $K$ and the initial time $\tau_{0}$ are two parameters that one can adjust to fit the experimental data on central rapidity density of produced hadrons. We will use the ideal version of CLVisc with a parametrized equation of state (EoS) s95p-v1\cite{Huovinen:2009yb} to obtain the hydrodynamic evolution of the bulk medium in 200 events of heavy-ion collisions in each centrality  to simulate jet transport in each bin of the initial transverse momentum transfer $p_{T i}$. We set the width of the bin in the initial transverse momentum transfer to be $\Delta p_{T i}=10$ GeV/$c$ and generate 1000 sets of initial jet showers from PYTHIA 8 in each bin for each of the 200 hydro events. The total number of events of initial jet production for each centrality in each $p_{T i}$ bin is therefore $N_{\rm event}=200\times 1000$. This is also the total number of events in each $p_{Ti}$ bin in p+p collisions.

The AMPT model employs the HIJING model \cite{Wang:1991hta,Gyulassy:1994ew} to generate the initial bulk parton or minijet production according to the Glauber model of nuclear collisions with the Woods-Saxon nuclear distribution. The geometrical distribution of the initial triggered jets in the transverse plane is sampled according to the initial minijet distribution in each AMPT event. The same AMPT event also provides the initial condition for the energy-momentum density distribution for CLVisc hydrodynamic simulations of the space-time evolution of the bulk medium in which jet transport is simulated according to the LBT model. The centrality classes of heavy-ion collisions are defined according to the initial parton multiplicity distribution and the averaged number of participant nucleons $\langle N_{\rm part}\rangle$ in each centrality class is computed accordingly. The interaction rate in Eq.~(\ref{eq-rate}) and jet transport coefficient in Eq.~(\ref{eq-qhat}) are all proportional to the medium parton density which will vanish in the hadronic phase of the bulk medium. The jet-medium interaction will be terminated in the hadronic phase and the final partons will be used for jet reconstruction within the FASTJET.  Equation (\ref{eq-jetcrs}) will then be used to calculate the differential single inclusive jet cross section per binary nucleon-nucleon pair in heavy-ion collisions within a given centrality class. The suppression factor $R_{\rm AA}(p_T)$ is defined [Eq.~(\ref{eq:raa})]  as the ratio between this cross section per binary nucleon-nucleon pair in heavy-ion collisions and that of single inclusive jet cross section in p+p collisions which is calculated from the same PYTHIA 8 events that provide the initial jet shower configurations for simulations of jet transport within LBT.

\begin{figure}[htbp]
    \includegraphics[width=8.5cm]{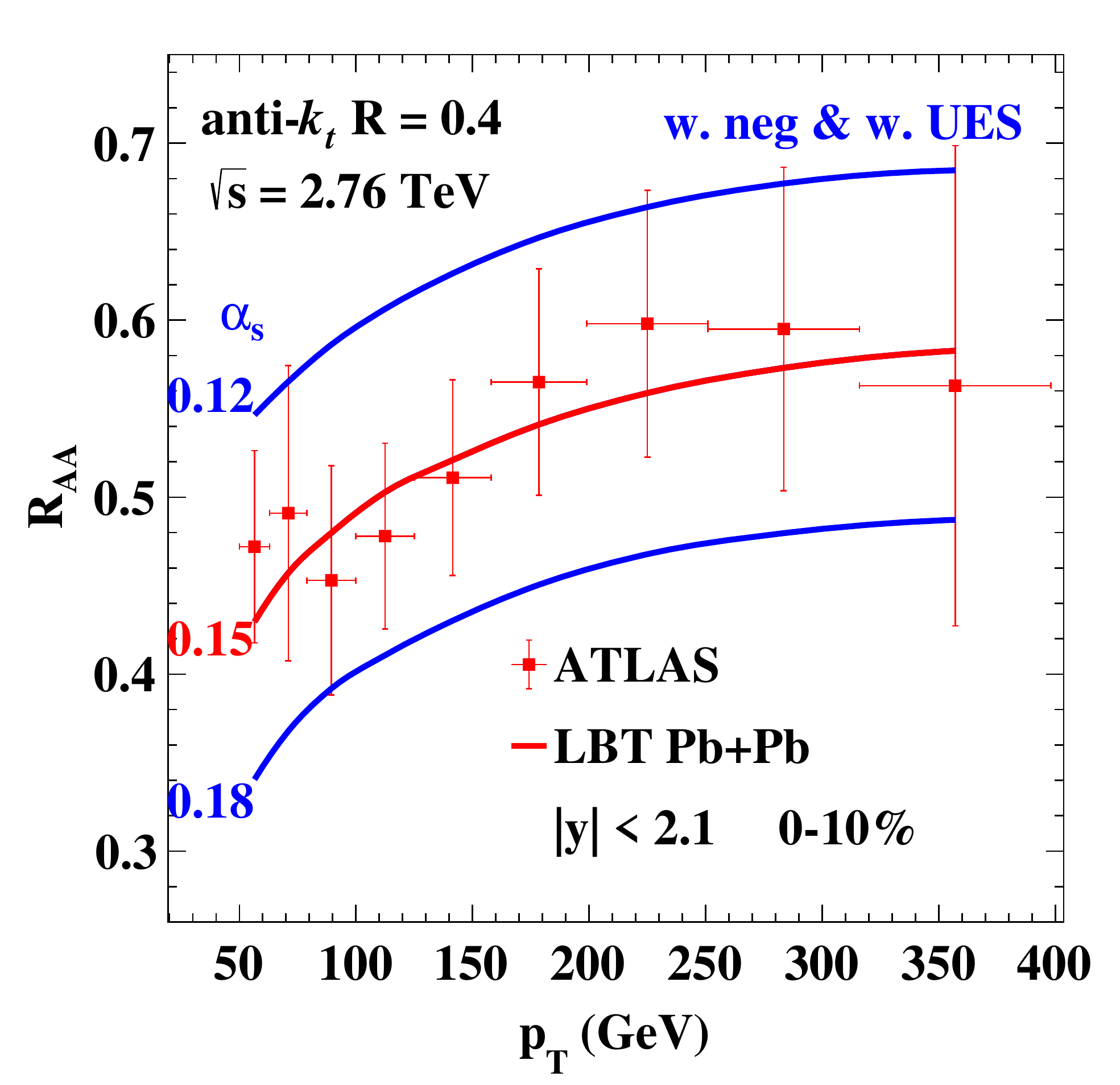}
    \caption{(Color online) The suppression factor $R_{\rm AA}$ of single inclusive jet spectra in the central rapidity $|y|<2.1$ region of 0-10\% central Pb+Pb collisions at $\sqrt{s}=2.76$ TeV from LBT simulations with different values of $\alpha_{\rm s}$ as compared to the ATLAS data at the LHC \cite{Aad:2014bxa}.  UES and ``negative" partons are both included in the jet reconstruction with $R=0.4$ and anti-$k_t$ jet-finding algorithm.} 
    \label{RAA_alphas}
\end{figure}

\begin{figure}[htbp]
    \includegraphics[width=8.5cm]{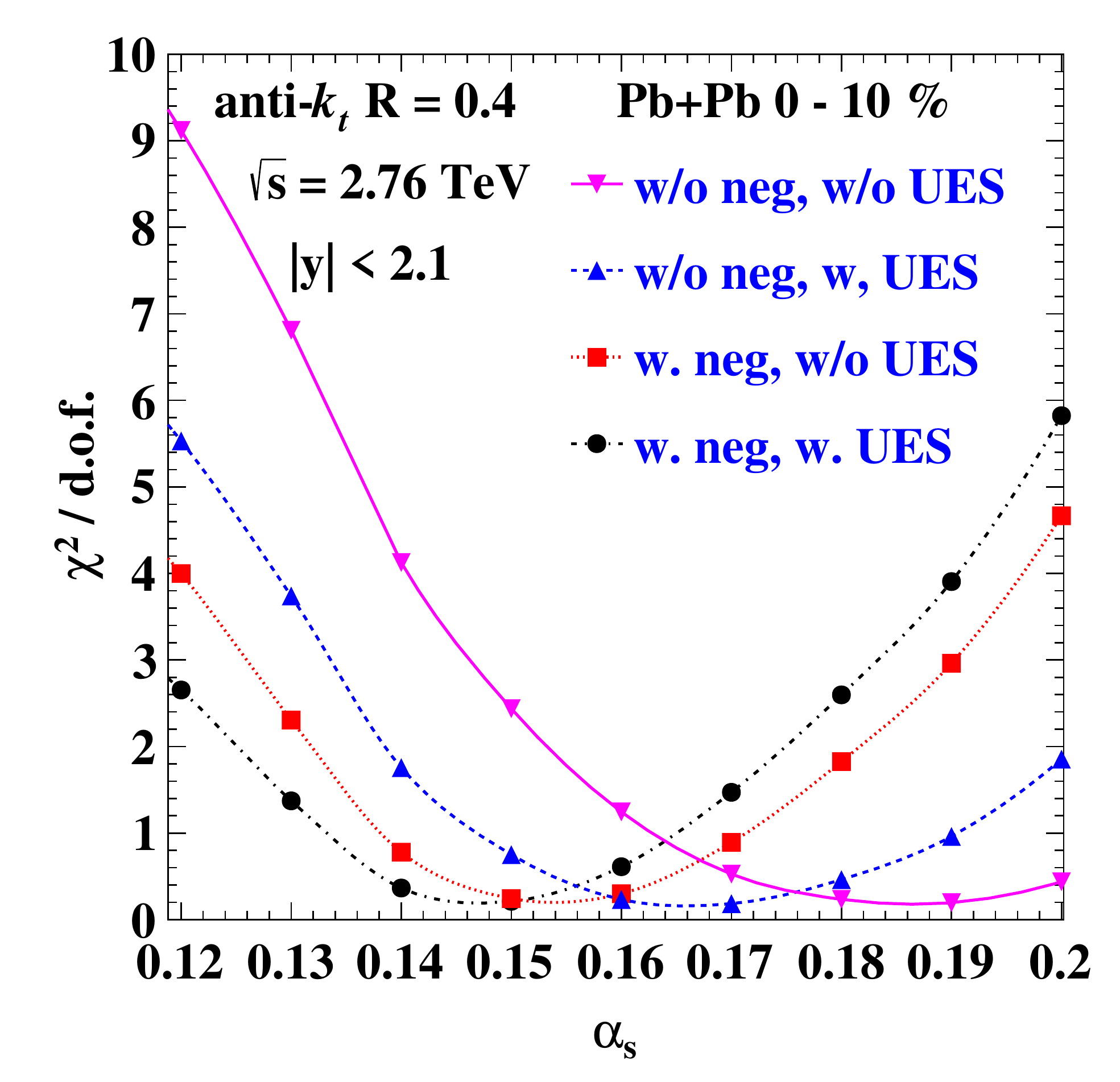}
    \caption{(Color online) $\chi^{2}$/d.o.f. of LBT fits to ATLAS data \cite{Aad:2014bxa} on $R_{\rm AA}(p_T)$ as a function of $\alpha_{s}$ in 0-10\% central Pb + Pb collisions at $\sqrt{s} = 2.76$ TeV with anti-$k_{t}$ algorithm and jet-cone size $R = 0.4$ in jet rapidity range $|y| < 2.1$, (black line with circle) with ``negative" partons and UES, (red line with square) with ``negative" partons but without UES, (blue line with uptriangle) with UES but without ``negative" partons, and (purple line with downtriangle) without ``negative" partons and UES.} 
    \label{chi010}
\end{figure}

\subsection{Suppression of single inclusive jet spectra}

Shown in  Fig.~\ref{RAA_alphas} are suppression factors for single inclusive jet production in the central rapidity $|y|<2.1$ region of 0-10\% central Pb+Pb collisions at $\sqrt{s}=2.76$ TeV from LBT simulations with different values of the fixed strong coupling constant $\alpha_{\rm s}$ as compared to the ATLAS data at the LHC \cite{Aad:2014bxa}.  Underlying event background subtraction (UES) and ``negative" partons due to back-reaction (diffusion wake) have both been included in the jet reconstruction and determination of the final jet transverse momentum using FASTJET with anti-$k_T$ algorithm and jet-cone size $R=0.4$. The central line is the LBT result with a value of $\alpha_{\rm s}=0.15$ that best fits the ATLAS data according to the $\chi^2$ distribution as shown in Fig.~\ref{chi010} in which we also show the $\chi^2$/d.o.f. (degrees of freedom) from fits of LBT results to the ATLAS data with different options on whether  ``negative" partons and UES are included in the jet reconstruction from LBT calculations. One can see from Fig.~\ref{chi010} that both ``negative" partons from the back-reaction and the UES have non-negligible effects on the reconstructed jet energy and the suppression factor for single inclusive jet spectra in heavy-ion collisions. Both effects reduce the transverse energy within the cone of the reconstructed jets.  These effects are more important for jets with large radii. 
The effect of UE is more important for low energy jets while the effect of ``negative" partons are non-negligible for jets at all energies.
With both effects included, one needs smaller interaction strength within the LBT model to fit the experimental data on single inclusive jet suppression in heavy-ion collisions.  They, however, do not change the minimum values of $\chi^2$/d.o.f. because of large uncertainties in the experimental data. With slightly different $\alpha_{\rm s}$, they can all describe the experimental data equally well. 

\begin{figure}[htbp]
    \includegraphics[width=8.5cm]{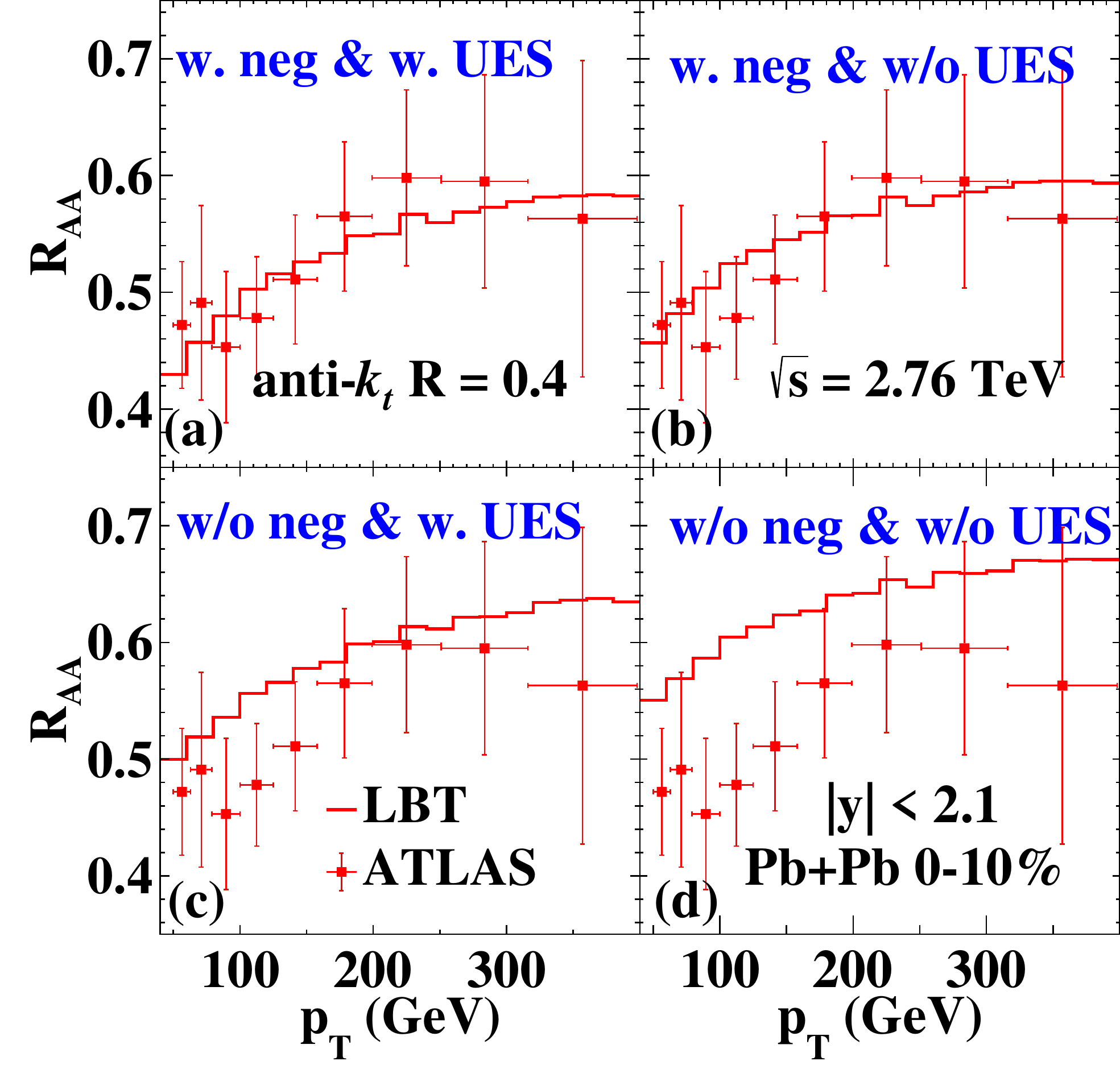}
    \caption{(Color online) The suppression factor $R_{\rm AA}$ of single inclusive jet spectra in the central rapidity $|y|<2.1$ region of 0-10\% central Pb+Pb collisions at $\sqrt{s}=2.76$ TeV from LBT simulations with fixed  $\alpha_{\rm s}=0.15$ as compared to the ATLAS data at the LHC \cite{Aad:2014bxa}.  
The jet reconstruction with $R=0.4$ and anti-$k_t$ algorithm includes four different options on ``negative" partons and UES: (a)  with both ``negative" partons and UES, (b) with ``negative" partons but without UES, (c) with UES but without ``negative" partons, and (d) without ``negative" partons and UES.} 
    \label{RAA_4opts}
\end{figure}

As another illustration of the effects of  ``negative" partons and UES on the single inclusive jet suppression, we show in Fig.~\ref{RAA_4opts} the suppression factors $R_{\rm AA}(p_T)$ for 0-10\% central Pb+Pb collisions at $\sqrt{s}=2.76$ TeV from LBT simulations with fixed $\alpha_{\rm s}=0.15$ and different options on  ``negative" partons and UES as compared to the ATLAS data.  Both effects lead to a bigger jet energy loss and therefore smaller values of the suppression factor, though the effect of ``negative" partons is larger. Without ``negative" partons, the effect of UES is also understandably larger than with  ``negative" partons. One can also see this from the $\chi^2$/d.o.f. distribution in Fig.~\ref{RAA_alphas} by comparing the effects of UES when ``negative" partons are included or not.
We will examine the effect of  ``negative" and recoil partons on the jet energy loss in more detail in the next section. 

We note that the fixed value of $\alpha_{\rm s}=0.15$ from the best fits to experimental data is only an effective strong coupling constant in the elastic scattering matrix elements and radiative gluon spectra in the LBT model in which we use the perturbative Debye screening mass in Eq.~(\ref{eq-mud}) to regularize the collinear divergence. It is possible that other non-perturbative physics such as chromo-magnetic monopoles can play a role in the parton-medium interaction \cite{Liao:2008jg,Liao:2008dk,Xu:2015bbz,Xu:2014tda} that can effectively increase the screen mass.  Furthermore, the non-zero mass of thermal partons can also reduce the effective thermal parton density significantly in the interaction rate. These can both increase the value of the effective strong coupling constant in LBT in order to fit the experimental data. In the remainder of this paper, we will use this value of fixed $\alpha_{\rm s}$ for all LBT calculations that include both ``negative" partons and UES, unless otherwise specified.

\begin{figure}[htbp]
        \centering
        \includegraphics[width=8.5cm]{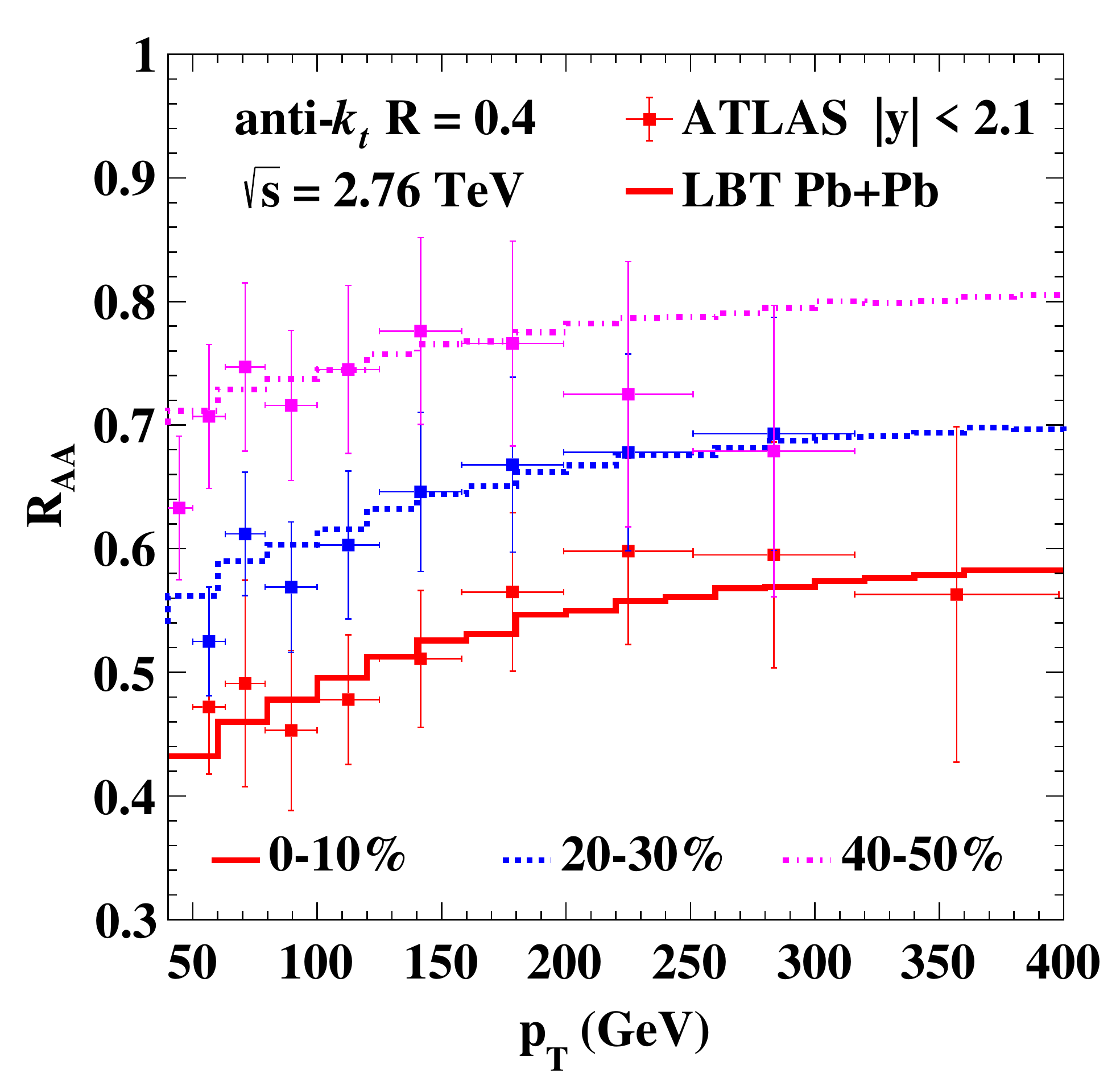}
       \caption{(Color online) LBT results on $R_{\rm AA}(p_T)$ in the central rapidity $|y|<2.1$ region of Pb+Pb collisions at $\sqrt{s}=2.76$ TeV for different centralities as compared to ATLAS data~\cite{Aad:2014bxa}.}
    \label{RAA_ctrl}
\end{figure}

\begin{figure}[htbp]
        \centering
        \includegraphics[width=8.5cm]{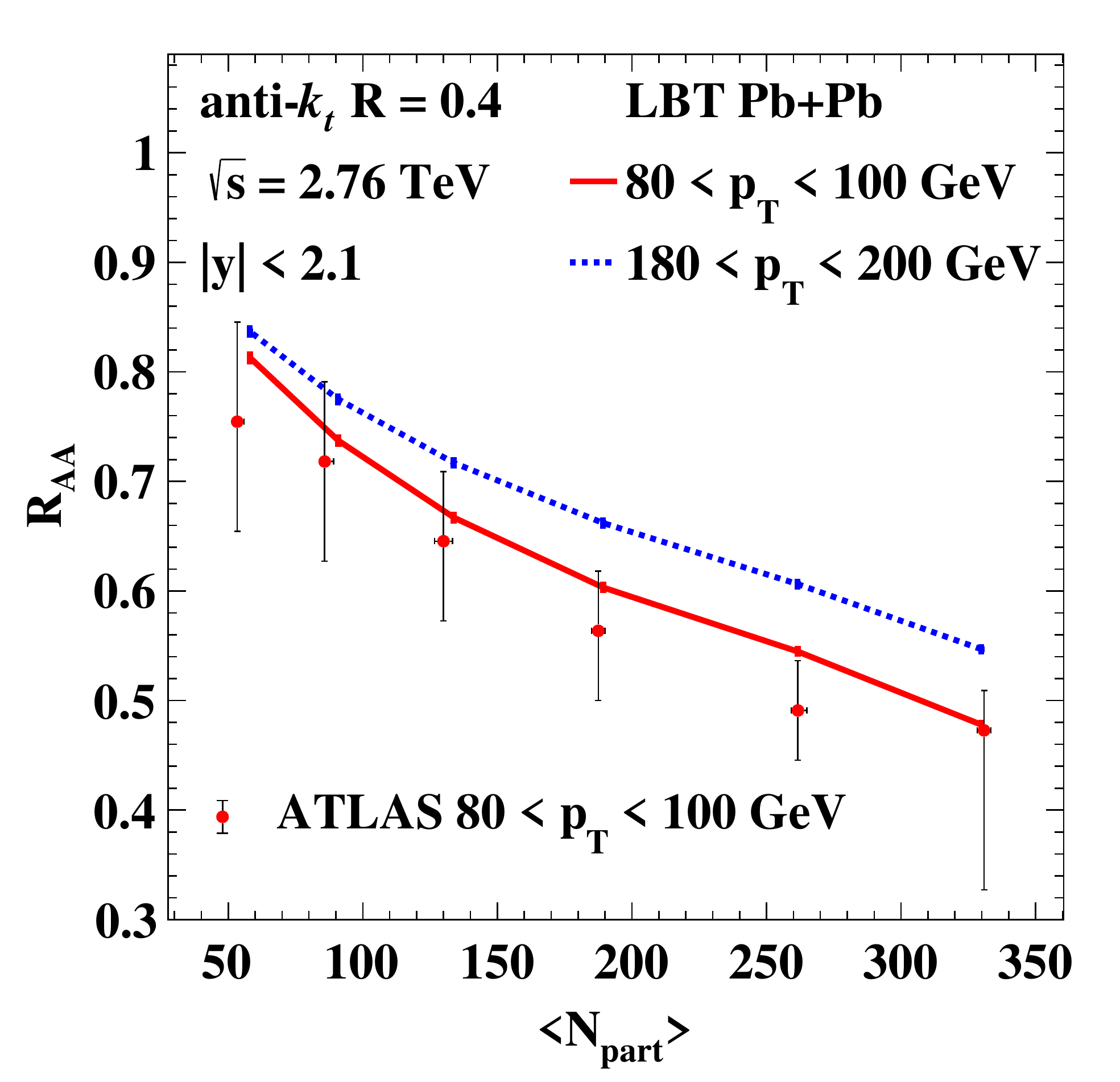}
       \caption{(Color online) LBT results on $R_{\rm AA}$ in Pb+Pb collisions at $\sqrt{s}=2.76$ TeV as a function of the number of nucleon participants $\langle N_{\rm part}\rangle$ in each centrality bin in two $p_T$ ranges, $p_T=80-100$ (solid), $180-200$ GeV/$c$ (dashed), as compared to experimental data from ATLAS~\cite{Aad:2014bxa}.}
    \label{RAA_Npart}
\end{figure}

With the only adjustable parameter $\alpha_{\rm s}$ fixed through the best fit to the ATLAS data on single inclusive jet suppression in 0-10\%  central Pb+Pb collisions at $\sqrt{s}=2.76$ TeV, we can predict the suppression factors for other centralities, rapidities and colliding energies. Shown in Fig.~\ref{RAA_ctrl}
are suppression factors for single inclusive jet spectra in three different centrality bins of Pb+Pb collisions at $\sqrt{s}=2.76$ TeV as compared to the ATLAS data. LBT results agree well with the data within the experimental errors. We have also calculated the inclusive jet suppression factor in 6 different centrality bins of Pb+Pb collisions at $\sqrt{s}=2.76$ TeV and plot it as a function of the mean number of participant nucleons $\langle N_{\rm part}\rangle$ in Fig.~\ref{RAA_Npart} for two different ranges of transverse momentum $p_T=80-120$ (solid line), $180-200$ GeV/$c$ (dashed line) as compared to ATLAS data at $p_T=80-120$ GeV/$c$. The LBT model can also describe well the experimental data on the centrality dependence of the single jet suppression.

\begin{figure}[htbp]
        \centering
        \includegraphics[width=8.5cm]{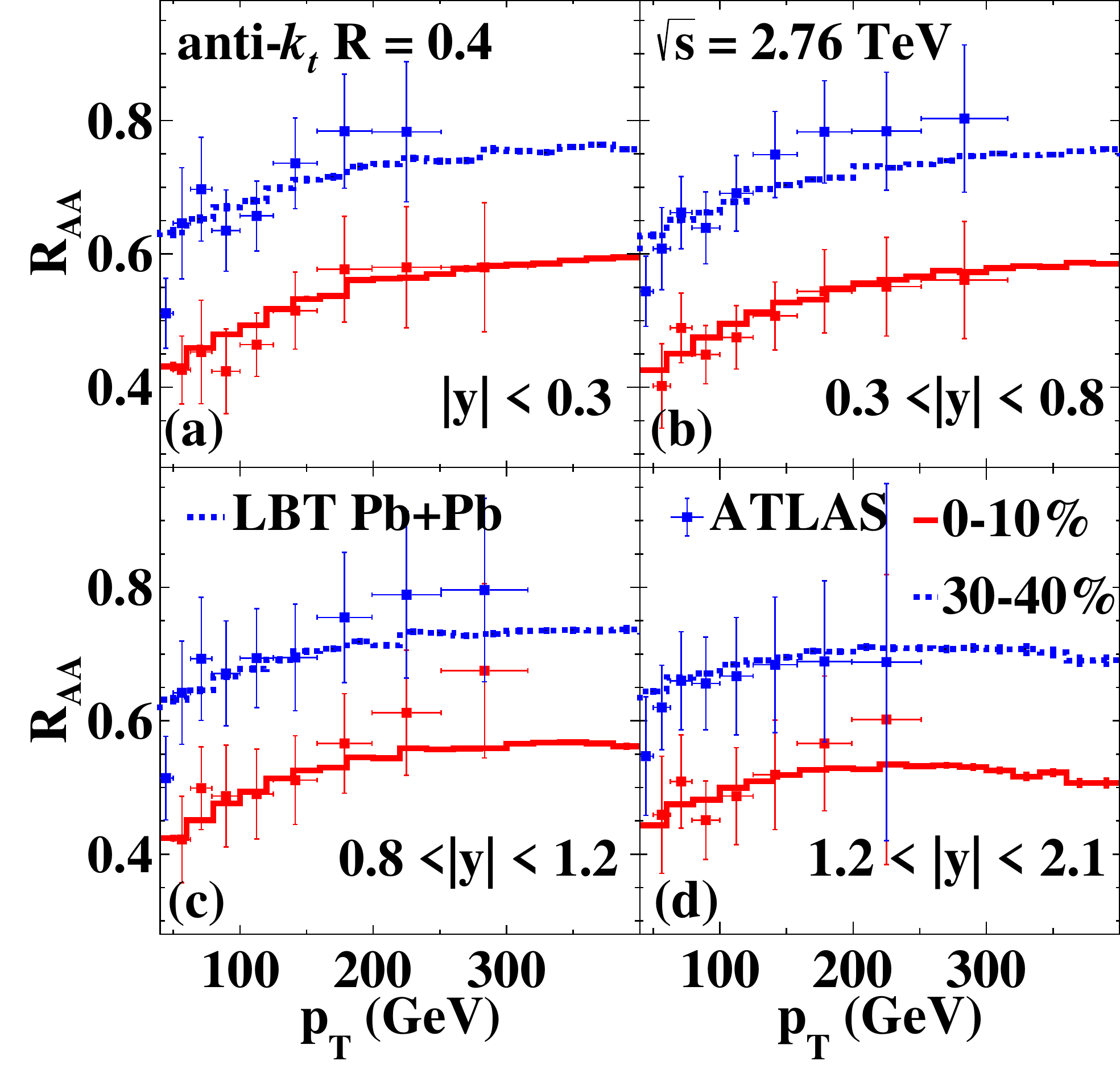}
    \caption{(Color online) LBT results on $R_{\rm AA}(p_T)$ in four different jet rapidities of (red solid) 0-10\% and (blue dashed) 30-40\% central Pb+Pb collisions at $\sqrt{s}=2.67$ TeV as compared to ATLAS data~\cite{Aad:2014bxa}.}
    \label{RAA_ctrl_rap}
\end{figure}

In Fig.~\ref{RAA_ctrl_rap}, we show the LBT results on single inclusive jet suppression factors in four different rapidity regions in 0-10\% central (solid lines) and 30-40\% semi-central (dashed lines) Pb+Pb collisions at $\sqrt{s}=2.76$ TeV. The suppression factor has a very weak rapidity dependence within $|y|<2.1$ consistent with ATLAS experimental data.

\begin{figure}[htbp]
    \includegraphics[width=8.5cm]{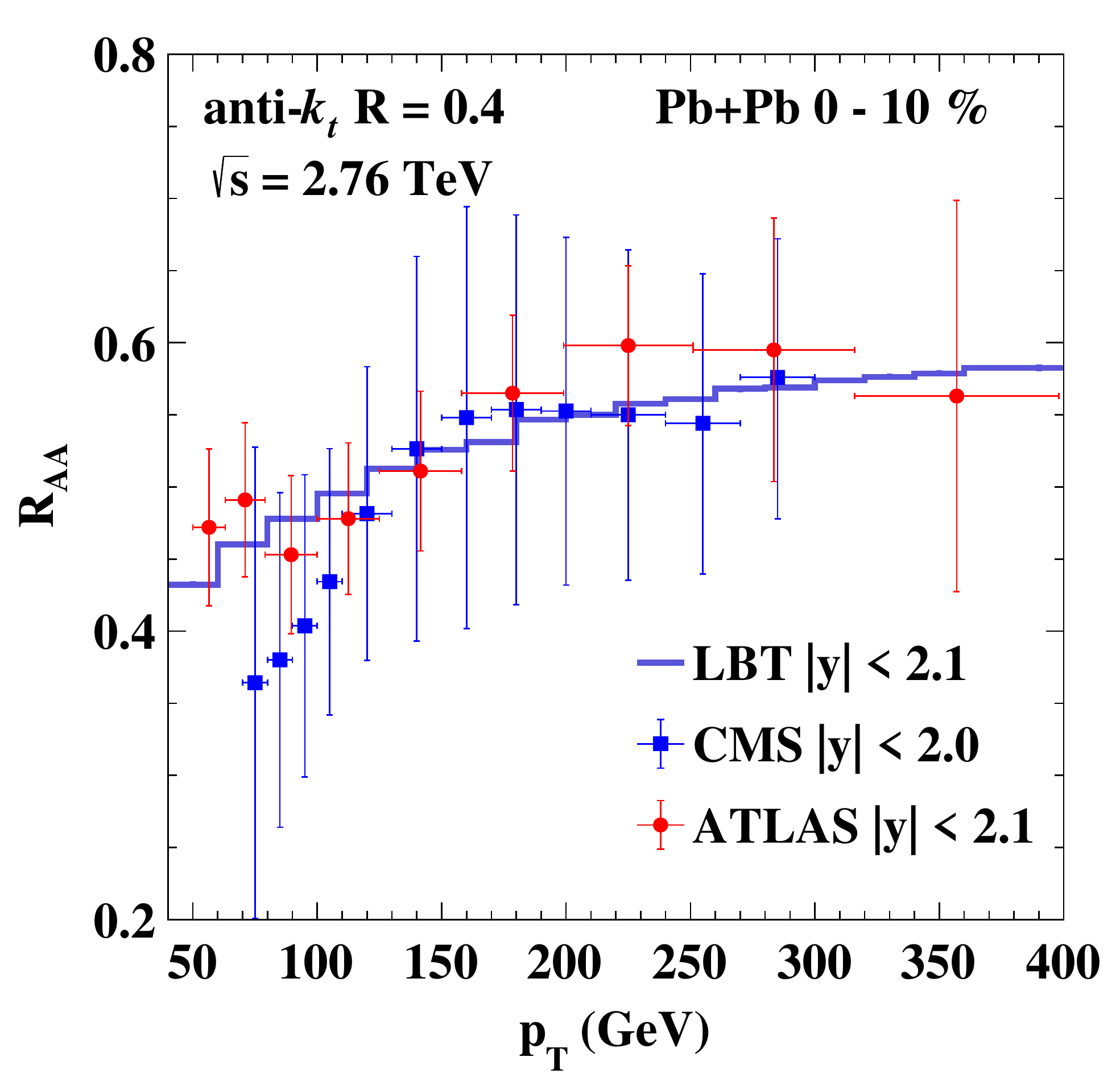}
    \caption{(Color online) Experimental data on $R_{\rm AA}(p_T)$ from ATLAS \cite{Aad:2014bxa} (red circle) and CMS \cite{Khachatryan:2016jfl} (blue square) for 0-10\% central Pb+Pb collisions at $\sqrt{s}=2.76$ TeV are compared to LBT calculations.}
    \label{RAA_Exp}
\end{figure}

LBT results on the single jet suppression factor in the central rapidity region of 0-10\% central Pb+Pb collisions at $\sqrt{s}=2.76$ are also compared to data from both ATLAS~\cite{Aad:2014bxa} and CMS \cite{Khachatryan:2016jfl} experiment at LHC in Fig.~\ref{RAA_Exp}. Data from both experiments are consistent with each other within their respective errors and with LBT calculations.

\section{Colliding energy and transverse momentum dependence of jet suppression}

\label{sec:jetsuppression}

\subsection{Colliding energy dependence}

In order to calculate the suppression of single inclusive jet spectra at different colliding energies, one first has to provide the initial conditions for the 3+1D hydrodynamic evolution. In our study here we use the initial parton production from the AMPT model for the initial condition for CLVisc hydrodynamic calculations. The scale factor in Eq.~(\ref{eq:Pmu}) is adjusted so that the final charged hadron rapidity density from the hydrodynamic calculation fits the experimental data in 0-10\% central Pb+Pb collisions at $\sqrt{s}=2.76$ and 5.02 TeV, respectively \cite{Abbas:2013bpa,Adam:2016ddh}. There is an increase of about 20\% in the charged hadron multiplicity density from 2.76 to 5.02 TeV. The corresponding event averaged initial temperature at the center of 0-10\% central Pb+Pb collisions is 469 and 529 MeV at an initial time $\tau_0=0.5$ fm/$c$, respectively, at these two colliding energies. 

\begin{figure}[htbp]
    \includegraphics[width=8.5cm]{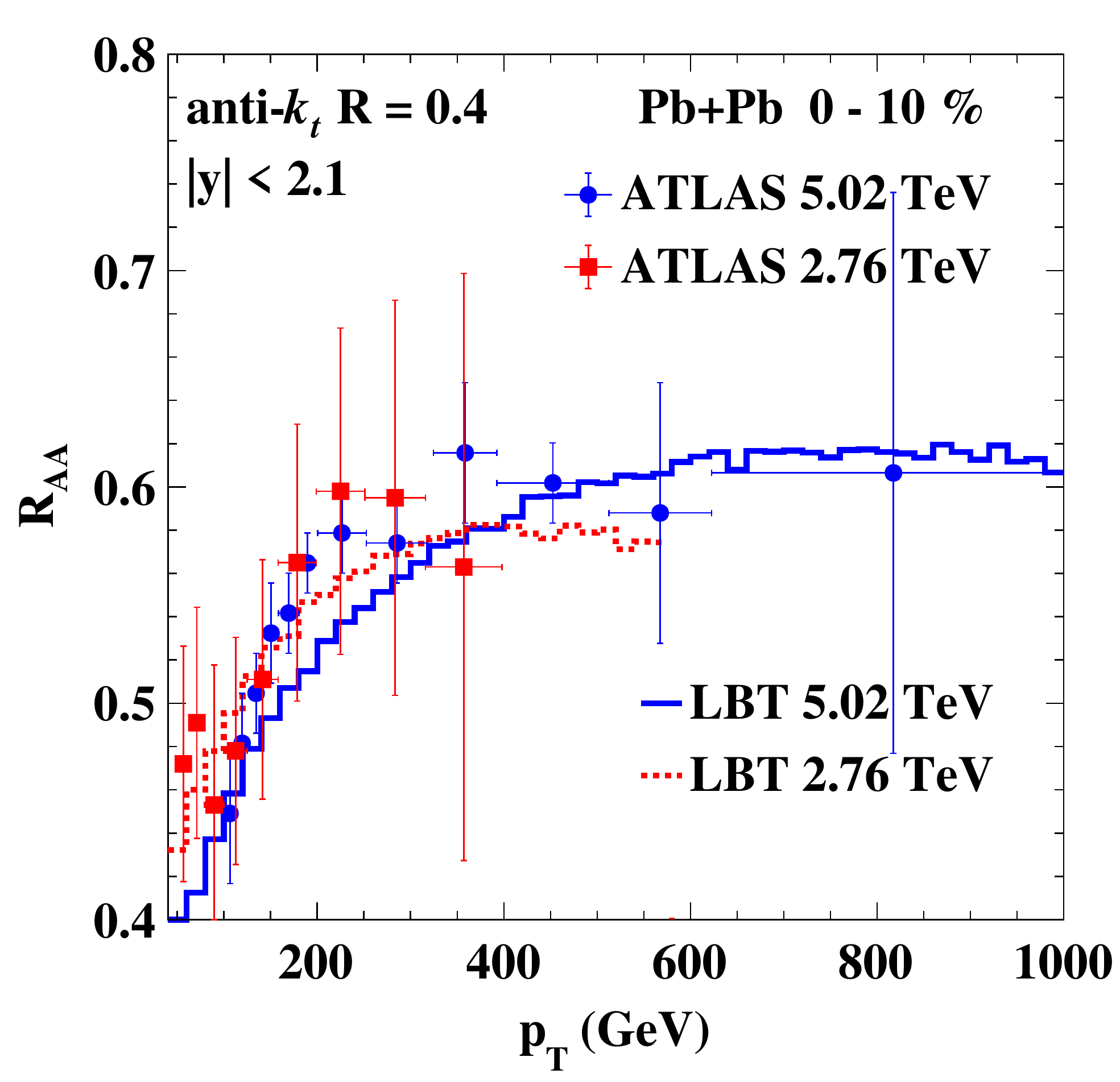}
    \caption{(Color online) LBT results on $R_{\rm AA}(p_T)$ in central rapidity $|y|<2.1$ for single inclusive jet spectra in 0-10\% central Pb+Pb collisions at $\sqrt{s} = 2.76$ (red dashed line) and 5.02 TeV (blue solid line) as compared to ATLAS data~\cite{Aad:2014bxa,Aaboud:2018twu}.}
    \label{RAA_twoEnergy}
\end{figure}

We assume the effective strong coupling constant in LBT is independent of the local temperature in this study and therefore can predict the suppression factor for single inclusive jet spectra in Pb+Pb collisions at $\sqrt{s}=5.02$ TeV as shown in  Fig.~\ref{RAA_twoEnergy} together with the latest data from ATLAS experiment \cite{Aad:2014bxa,Aaboud:2018twu}. One can observe two striking features in the LBT calculations which are consistent with the experimental data. The first feature is the very weak or none colliding energy dependence at LHC energy range despite the fact that the initial parton density at 5.02 TeV is about 20\% higher than at 2.76 TeV. The second feature is the weak transverse momentum dependence of the jet suppression factor in the range of the experimental coverage which is very different from the suppression factor for single inclusive charged hadrons \cite{Aamodt:2010jd,CMS:2012aa,Khachatryan:2016odn,Acharya:2018qsh}. 

\subsection{Jet energy loss distribution}

To understand the colliding energy and transverse momentum dependence of the jet suppression factor, we have to understand  the transverse momentum dependence of the average jet energy loss and its fluctuations. For given initial production point ${\bf r}$, impact parameter ${\bf b}$ and propagation direction $\phi_c$, we assume that the medium-modified single inclusive jet distribution is given by the convolution of the jet distribution in vacuum $dN^{\rm jet}_{(c)}(p_{Ti},p_T)/dydp_T$ and the jet energy loss distribution $w_c(\Delta p_T, p_{T}, {\bf r},{\bf b},\phi_c)$, 
\begin{eqnarray}
\frac{d\widetilde{N}^{\rm jet}_{(c)}(p_{Ti},p_T,{\bf r},{\bf b,}\phi_c)}{dydp_T}&=&\int d\Delta p_T \frac{d^2N^{\rm jet}_c(p_{Ti},p_T+\Delta p_T)} {dp_T dy}
 \nonumber \\
&&\hspace{-0.4in} \times w_c(\Delta p_T, p_{T}+\Delta p_T, {\bf r},{\bf b},\phi_c),
\end{eqnarray}
where we assume that the implicit dependence of jet energy loss distribution $w_c$ on the initial hard parton's transverse momentum $p_{Ti}$ is only through an explicit dependence on the final jet transverse momentum $p_T$ in vacuum. Averaging over the energy loss fluctuation due to distribution of the production point and the propagation direction, one can define the energy loss
distribution for a given centrality class of A+A collisions as
\begin{eqnarray}
W^{(c)}_{\rm AA}(\Delta p_T, p_{T})&=&\int d^2{\bf r} d^2{\bf b}  t_A(r) t_A(|{\bf b}-{\bf r}|) \frac{d\phi_c}{2\pi} \nonumber \\
&\times& \frac{w_c(\Delta p_T, p_{T}, {\bf r},{\bf b},\phi_c)}{\int d^2{\bf r }d^2{\bf b}  t_A(r) t_A(|{\bf b}-{\bf r}|)}.
\end{eqnarray}
The cross section for single inclusive jet production in A+A collision in Eq.~(\ref{eq:cs.aa}) can be rewritten as
\begin{eqnarray}
\frac{d \sigma^{\rm jet}_{\rm AA}}{dp_{T}dy} & = & \int dp_{Ti} d\Delta p_T \frac{d \sigma^{{\rm AA}(c)}_{\rm LO}}{dp_{Ti}} W^{(c)}_{\rm AA}(\Delta p_T, p_{T}+\Delta p_T)
\nonumber \\
 &\times& \frac{d^2N^{\rm jet}_c(p_{Ti},p_T+\Delta p_T)} {dp_T dy},
\label{eq:jetaa}
\end{eqnarray}
where the effective LO pQCD jet production cross section per binary nucleon-nucleon interaction is defined as
\begin{eqnarray}
\frac{d \sigma^{{\rm AA}(c)}_{\rm LO}}{dp_{Ti}} & = & 2 p_{Ti}\sum_{a,b,d}  \int dy_c dy_d  x_a f_{a/A} (x_a, \mu^2)
 \nonumber\\ 
 & &  \times x_b f_{b/A} (x_b, \mu^2)  \frac{d\hat\sigma_{ab\to cd}}{dt}.
\label{eq:LOaa}
\end{eqnarray}
If we neglect the small nuclear modification of parton distribution functions at very large momentum scale~\cite{Eskola:2009uj,Ru:2016wfx}, $d \sigma^{{\rm AA}(c)}_{\rm LO}/dp_{Ti}\approx d \sigma^{{\rm pp}(c)}_{\rm LO}/dp_{Ti}$, the modification factor for single inclusive jet production in A+A collisions can be written as
\begin{eqnarray}
 R_{\rm AA} (p_{T}) &\approx& \int d\Delta p_T W_{\rm AA}(\Delta p_T, p_T+\Delta p_T) \nonumber \\
 &\times& \frac{d\sigma^{\rm jet}_{\rm p+p}(p_{T} + \Delta p_T)}{d\sigma^{\rm jet}_{\rm p+p}(p_{T})},
 \label{shift}
\end{eqnarray}
where $W_{\rm AA}$ is the flavor-averaged parton energy loss distribution for a given centrality class of A+A collisions and jet-cone size $R$. If the average jet energy loss is small, the above jet suppression factor can be approximated with
\begin{equation}
 R_{\rm AA} (p_{T}) \approx \frac{d\sigma^{\rm jet}_{\rm p+p}(p_{T} + \langle \Delta p_T\rangle)}{d\sigma^{\rm jet}_{\rm p+p}(p_{T})},
 \label{shift2}
\end{equation}
where the average jet energy loss is given by
\begin{equation}
\langle \Delta p_T\rangle(p_T)=\int d\Delta p_T \Delta p_T W_{\rm AA}(\Delta p_T, p_T),
 \label{aveloss}
\end{equation}
which should depend on the vacuum jet energy $p_T$, colliding energy $\sqrt{s}$, centrality and the jet-cone size $R$.

\begin{figure}[htbp]
    \centering
    \includegraphics[width=8.5cm]{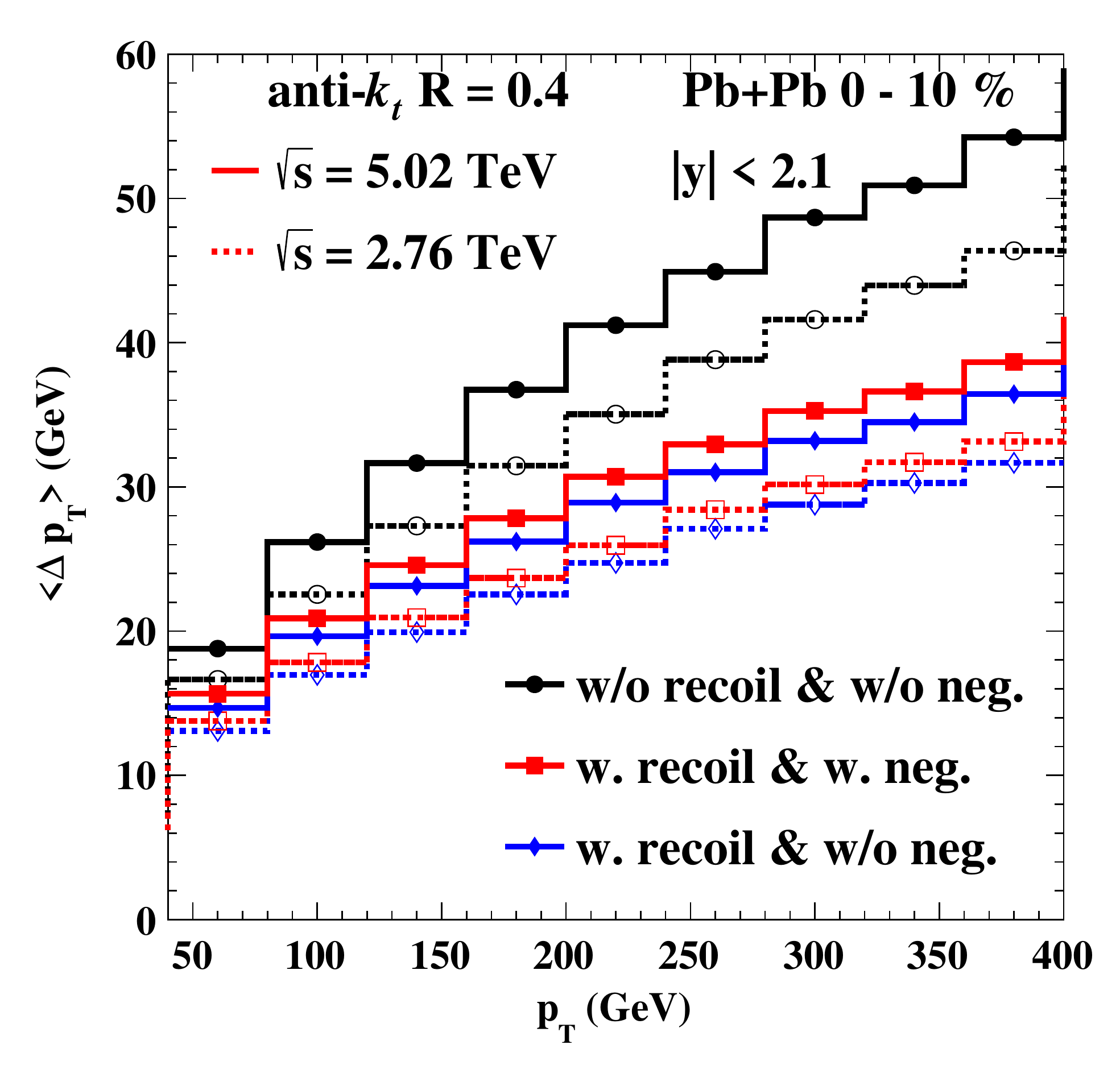}
    \caption{(Color online) Average jet transverse energy loss as a function of vacuum jet $p_{T}$ with anti-$k_t$ and  $R = 0.4$ 
    in $|y| < 2.1$ of central 0 - 10 \% Pb+Pb collisions  at (solid) $\sqrt{s} = 5.02$ GeV  and  (dash) 2.76 TeV. Black lines with circles are the LBT results without recoil and ``negative" partons, while red lines with squares are with recoil and ``negative" partons and blue lines with diamonds are with recoil but without ``negative" partons.}
    \label{pTloss_twoEnergy}
\end{figure}

To illustrate the colliding energy and transverse momentum dependence of the jet energy loss and its fluctuation, we first show the averaged energy 
loss $\langle \Delta p_T\rangle$ in Fig.~\ref{pTloss_twoEnergy} for leading jets in the 0-10\% central Pb+Pb collisions at two colliding energies, $\sqrt{s}=2.76$ and 5.02 TeV, from LBT simulations. In the calculations, the leading jet with a large cone size $R=1$ from PYTHIA 8 in each event and the associated jet shower partons are identified. These jet shower partons are then used for the reconstruction of the vacuum leading jet in p+p collisions with a given jet-cone size $R$ and UES. These same jet shower partons are allowed to propagation through the hydrodynamic medium in LBT and the transverse energy of the final medium-modified leading jet with cone size $R$ is calculated with the same jet-finding algorithm and UES. The difference between the final transverse energies of the vacuum and medium-modified leading jet is defined as the jet transverse energy loss as shown in Fig.~\ref{pTloss_twoEnergy} as a function of the vacuum jet transverse energy. An alternative definition of the jet energy loss is the energy difference between the leading jet in p+p and the leading jet in A+A in the same direction of the vacuum leading jet with the angular difference smaller than the jet-cone size, $\Delta r<R$.  The results are approximately the same.
The transverse jet energy loss at $\sqrt{s}=5.02$ TeV is indeed about 15\% larger than at $\sqrt{s}=2.76$ TeV in the $p_T=50-400$ GeV/$c$ range when the medium response (recoil and ``negative" partons) is taken into account in the calculation of the transverse energy of the medium-modified leading jet.  It increases with the vacuum jet transverse energy logarithmically similar to that of a single parton~\cite{Guo:2000nz,Wang:2001ifa}.  As we will discuss later in detail, such a weak $p_T$-dependence of the jet transverse energy loss is caused by a combination of effects due to jet-induced medium response, radial expansion and jet flavor (quarks and gluons) composition.

\begin{figure}[htbp]
    \centering
    \includegraphics[width=8.5cm]{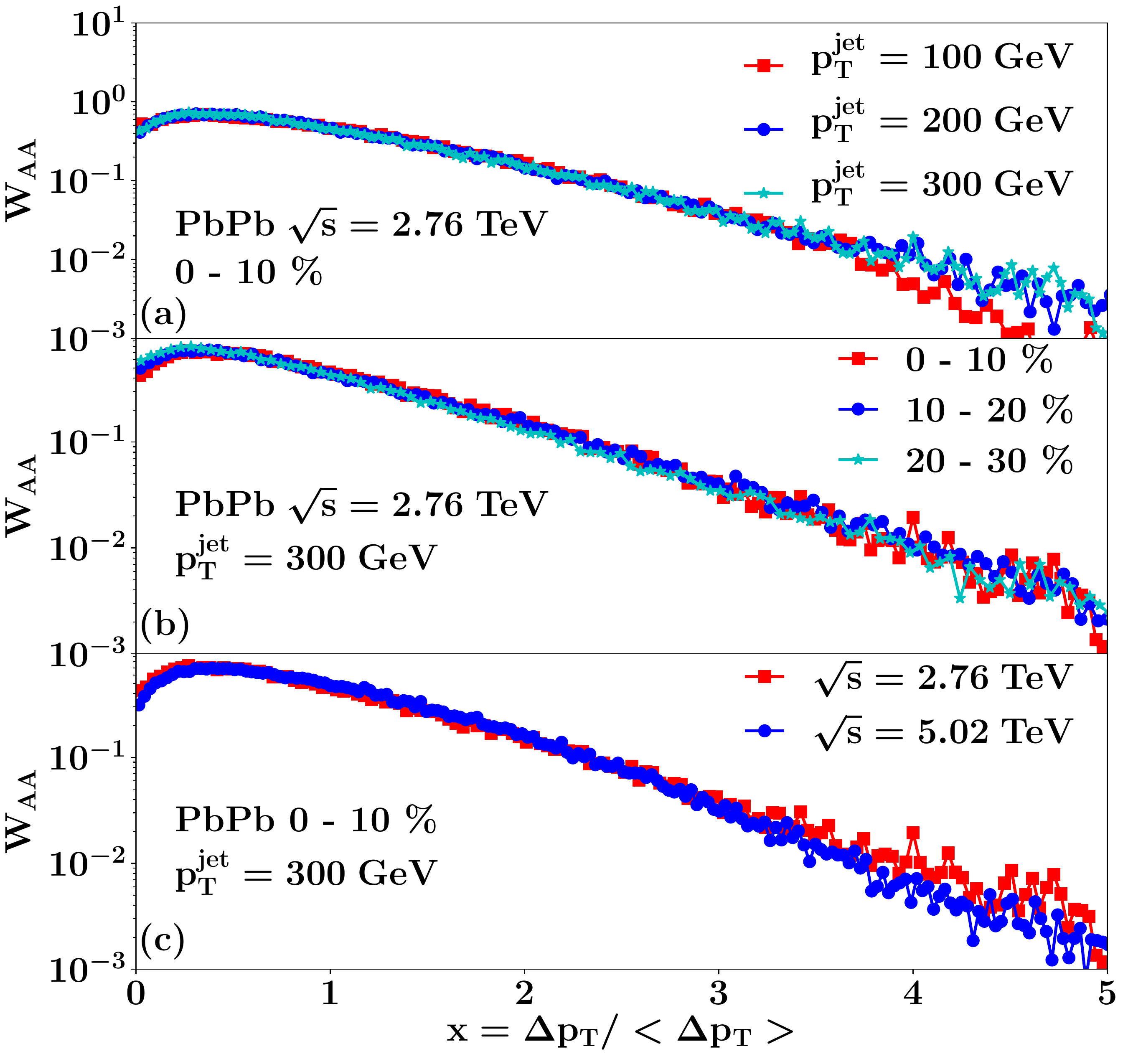}
    \caption{(Color online) LBT results on jet energy loss distribution $W_{\rm AA}(x)$ as a function of the scaled jet energy loss $x=\Delta p_T/\langle \Delta p_T\rangle$ in Pb+Pb collisions (a) for three different vacuum jet energies, (b) three different centralities and (c) two different colliding energies at LHC.}
    \label{elossdistr}
\end{figure}

We also show the jet energy loss distributions $W_{\rm AA}(\Delta p_T, p_T)$ as a function of the scaled variable $x=\Delta p_T/\langle \Delta p_T\rangle$ from LBT simulations in Fig.~\ref{elossdistr} for leading jets (a) with vacuum transverse momentum $p_T=100, 200, 300$ GeV/$c$ in 0-10\% central Pb+Pb collisions  at $\sqrt{s}=2.76$ TeV, (b) for $p_T=300$ GeV/$c$ in Pb+Pb collisions with different centralities (0-10\%, 10-20\%, 20-30\%) at $\sqrt{s}=2.76$ TeV,  and (c) for $p_T=300$ GeV/$c$ in 0-10\% Pb+Pb collisions at both $\sqrt{s}=2.76$ and 5.02 TeV. We can see that the jet energy loss distribution has an scaling behavior in the scaled variable $x=\Delta p_T/\langle \Delta p_T\rangle$ approximately independent of the vacuum jet $p_T$ and the colliding energy for a given centrality of heavy-ion collisions. The dependence of the jet energy loss distribution on the vacuum jet energy and colliding energy is only implicit through the average jet energy loss $\langle \Delta p_T\rangle(p_T, \sqrt{s})$. Such a scaling property of the jet energy loss distribution is essentially determined by the fluctuation of the jet energy loss caused by a scattering that can transport jet shower partons to the outside of the jet cone and the average number of such out-of-cone scatterings in a given centrality class of A+A collisions. It can be used to extract the jet energy loss distributions from experimental data on jet spectra in p+p and A+A collisions using the convolution relationship in Eq.~(\ref{shift}) \cite{He:2018gks}. Note that the scaling behavior of $W_{\rm AA}(x)$ will be violated at very large values of $x$ for finite values of the vacuum jet transverse momentum $p_T$ due to energy-momentum conservation since the total jet energy loss is limited by the initial or vacuum jet energy. This violation will only influence the tails of the scaling jet energy loss distributions as seen in Fig.~\ref{elossdistr} where the total jet energy loss is large.

\subsection{Understanding the colliding energy and transverse momentum dependence}

Given the jet energy loss distribution, $p_T$ and $\sqrt{s}$ dependence of the average jet transverse energy loss, one should be able to estimate the suppression of jet spectra by shifting jet production cross section as measured in p+p collisions through Eq.~(\ref{shift}) or (\ref{shift2}). As we can see in Fig.~\ref{jetCS_twoEnergy},  the shape of the single inclusive jet spectra at $\sqrt{s}=5.02$ TeV is much flatter than that at 2.76 TeV in the same $p_T$ range. This colliding energy dependence of the single inclusive jet spectra in p+p collisions is one of the deciding factors that will influence the energy and transverse momentum dependence of the jet suppression factor $R_{\rm AA}(p_T)$. 

\begin{figure}[htbp]
    \includegraphics[width=8.5cm]{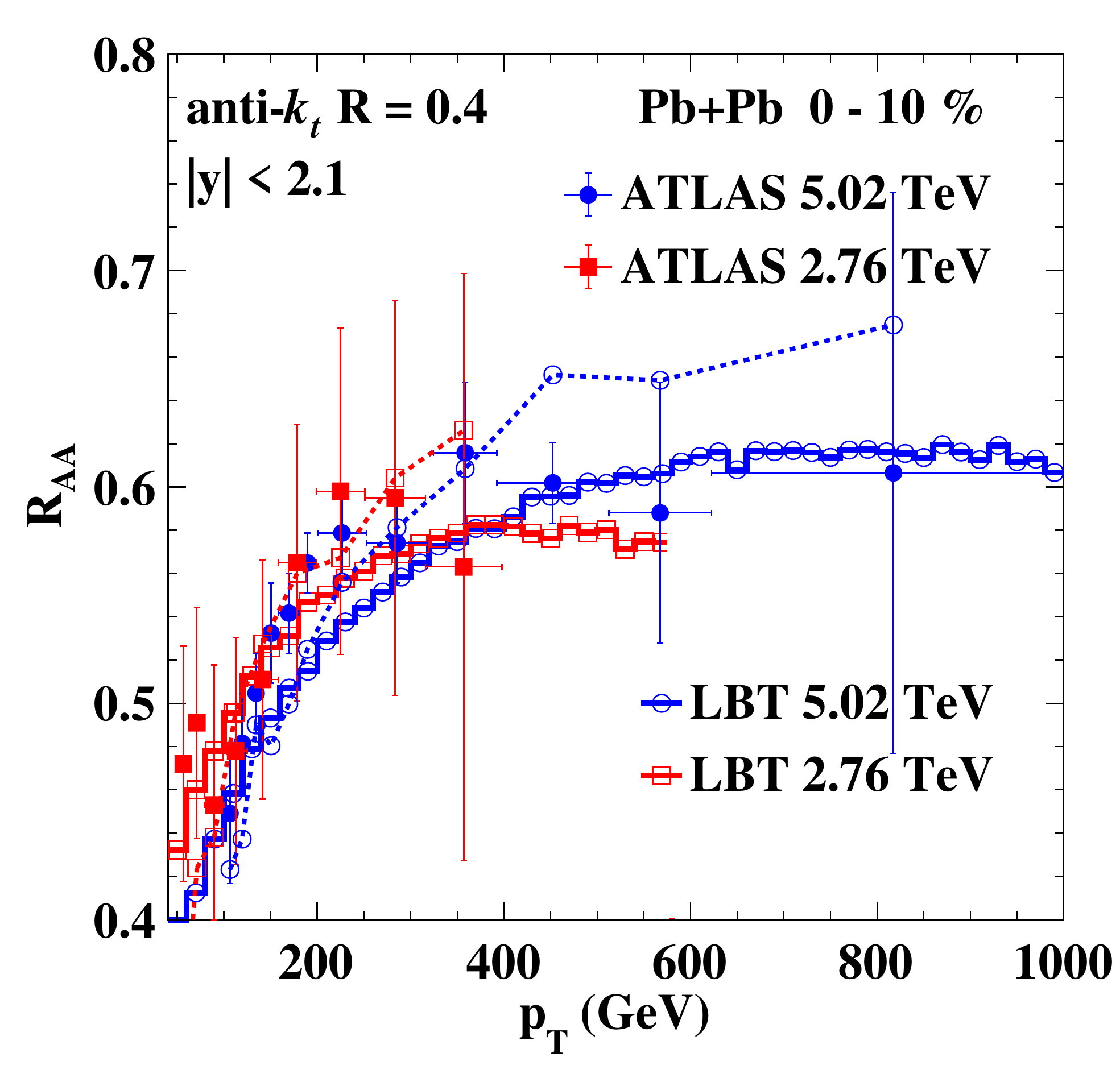}
    \caption{(Color online) Experimental data on $R_{\rm AA}$ for 0-10\% central Pb+Pb collisions at  (red solid squares)  $\sqrt{s} = 2.76$ TeV and  (blue solid circles) 5.02 TeV \cite{Aad:2014bxa,Aaboud:2018twu} as compared to  (solid lines)  LBT calculations and  (dashed) the suppression factor obtained by shifting the jet spectra in p+p collisions by the average jet energy loss from Fig.~\ref{pTloss_twoEnergy} according to Eq.~(\ref{shift2}).}
    \label{RAA_shift_twoEnergy}
\end{figure}

Shown in Fig.~\ref{RAA_shift_twoEnergy} are the jet suppression factors (dashed lines) obtained by shifting the transverse momentum in the jet production cross section in p+p collisions with the average transverse energy loss as shown in Fig.~\ref{pTloss_twoEnergy} according to Eq.~(\ref{shift2}), together with the full LBT calculations (solid lines) and ATLAS data. A scaling factor of $1.174$ and $1.165$ is multiplied to the shifted spectra at  $\sqrt{s} = 2.76$ and 5.02 TeV, respectively, to keep the number of inclusive jets the same. One can see that the colliding energy and the transverse momentum dependence of the jet suppression factor can be approximately determined by the behavior of the transverse energy loss and the shape of the initial jet production spectra. The approximate 15\% increase in the transverse energy loss from $\sqrt{s}=2.76$ to 5.02 TeV, as shown in Fig.~\ref{pTloss_twoEnergy}, is mostly offset by the decrease of the slope of the jet $p_T$ spectra (becoming flatter), leading to a suppression factor that has a very weak colliding energy dependence.  The initial jet production spectra in the large $p_T$ region at both colliding energies are more exponential than power-law-like in the large $p_T$ region due to the fall-off of parton distribution functions in the large momentum-fraction region. This shape of the initial production spectra coupled with the weak $p_T$-dependence of the transverse energy loss in these regions of $p_T$  leads to a very weak $p_T$ dependence of the jet suppression factor.  Note that the weak $p_T$ dependence of the jet transverse energy loss is partially caused by the influence of jet-induced medium response on the jet energy within a given cone size $R$ as shown in Fig.~\ref{pTloss_twoEnergy}.  A detailed analysis of the colliding energy and $p_T$ dependence of the suppression factor given the initial jet spectra in p+p collisions can provide important information about the jet energy loss distributions according to Eq.~(\ref{shift}). This has been investigated in detail in a separate study \cite{He:2018gks}.

\section{Effects of medium response, radial expansion and jet flavor}

\label{sec:eloss}

As we have shown in the previous section, the behavior of the suppression factor for single inclusive jets is closely related to the colliding energy and transverse momentum dependence of the jet energy loss due to jet-medium interaction in an expanding QGP. We will examine in this section the effects of medium response, radial expansion and jet flavor on the jet energy loss in detail. 

\subsection{Effects of medium response and radial expansion}

Similar to the calculation of jet energy loss in the last section, we focus on the leading jet in both p+p and central (0-10\%) Pb+Pb collisions. Only the jet shower partons associated with the leading jet within a large jet-cone size $R=1$ in PYTHIA 8 simulations of p+p collisions are used for propagation within LBT in 200 events of hydrodynamic profiles with fluctuating initial conditions for 0-10\% central Pb+Pb collisions. FASTJET is used to calculate the transverse energy of the vacuum and medium-modified leading jet with UE subtraction and the transverse energy loss is calculated for different jet-cone sizes. We choose three different jet-cone sizes $R=0.3$, 0.4 and 0.5 to investigate the dependence on the jet-cone size. To study the effect of radial expansion, we also compare to the case where the same jet shower partons propagate in a static medium with a constant temperature $T=0.28$ GeV and finite length (or propagation time)  $L=4$ fm. The length is approximately the average propagation length in 0-10\% central Pb+Pb collisions and the temperature is chosen such that the jet transverse energy loss for $R=0.4$ in the static medium is the same as that of a dynamically evolving medium in 0-10\% central Pb+Pb collisions at $\sqrt{s}=2.76$ TeV in the lowest $p_T$ bin in our study here.

\begin{figure}[htbp]
    \includegraphics[width=8.5cm]{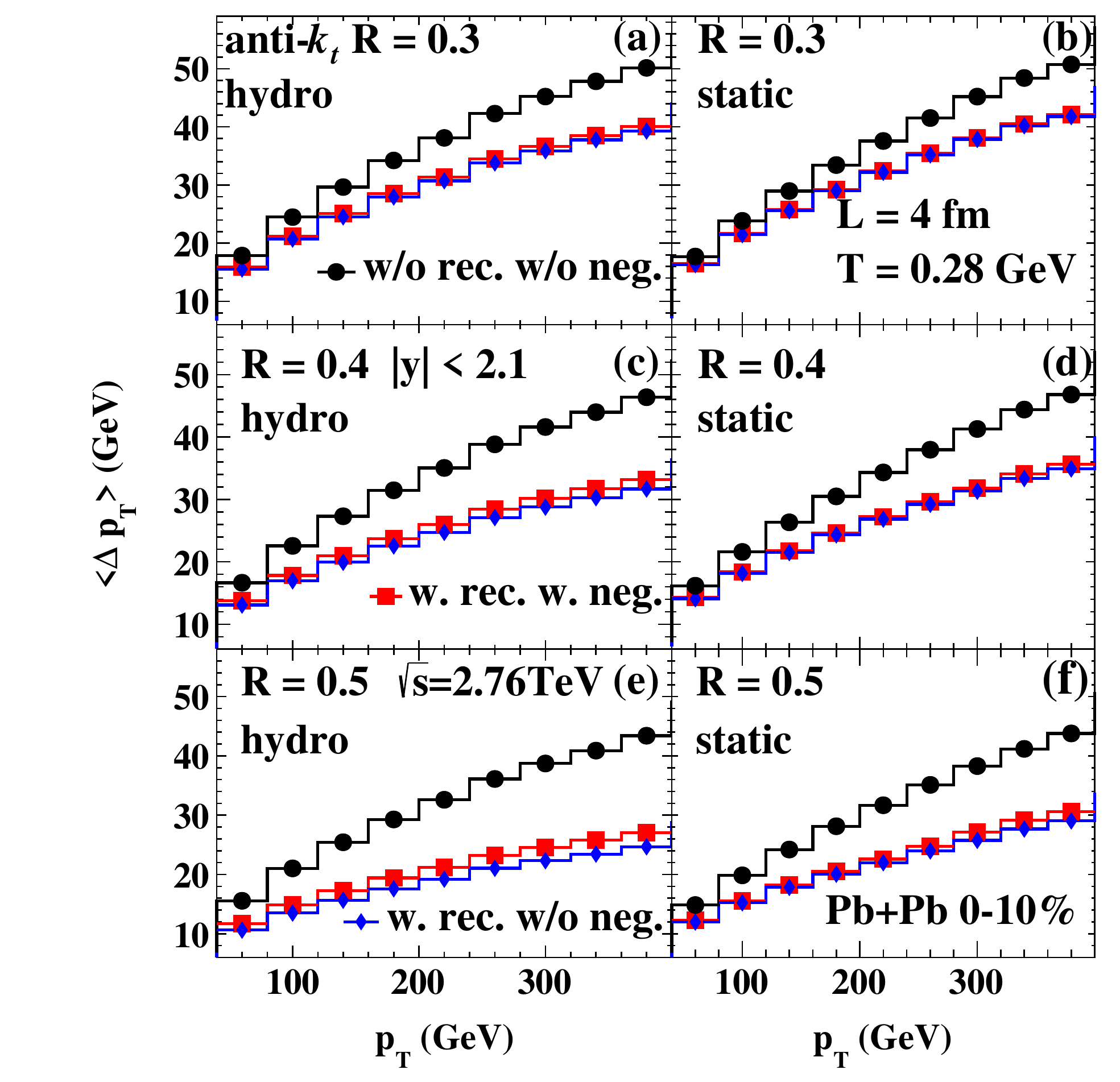}
    \caption{(Color online) LBT results on average $p_{T}$ loss $\langle \Delta p_T\rangle$ for jets in $|y| < 2.1$ as a function of the vacuum jet $p_{T}$ with anti-$k_t$ algorithm and $R = 0.3, 0.4, 0.5$ for [(a), (c), (e)] hydrodynamic background in central 0 - 10\% Pb+Pb collisions at $\sqrt{s} = 2.76$ TeV  and [(b), (d), (f)] static medium at $T = 0.28$ GeV with fixed length $L = 4$ fm. Black lines with circles are results without recoil and ``negative" partons, while red lines with squares are with recoil and  ``negative" partons and blue lines with diamonds are with recoil but without ``negative" partons.}
    \label{pTloss_2760}
\end{figure}
\begin{figure}[htbp]
    \includegraphics[width=8.5cm]{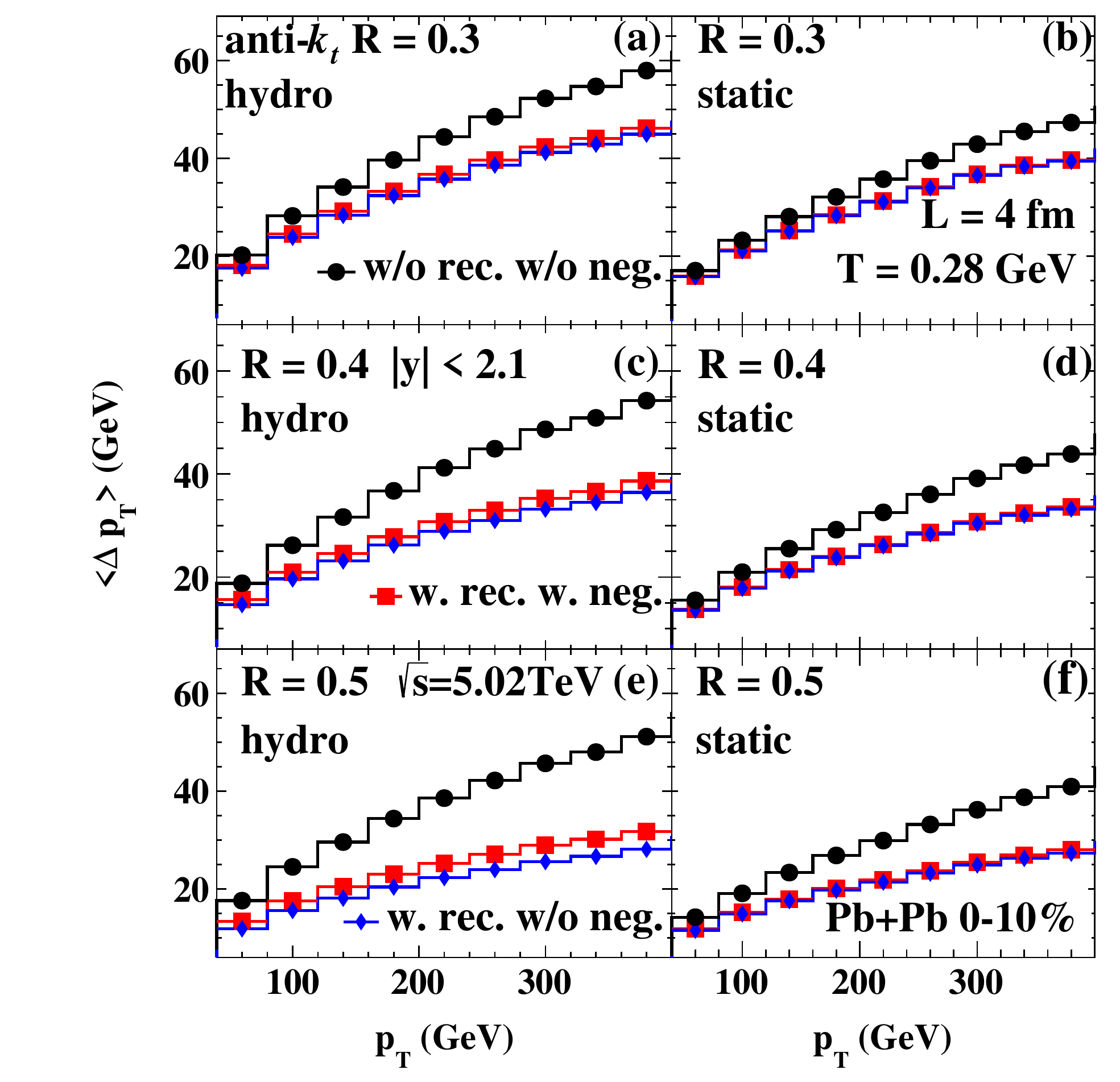}
    \caption{The same as Fig.~\ref{pTloss_2760} except for $\sqrt{s} = 5.02$ TeV.}
      \label{pTloss_5020}
\end{figure}

Shown in Figs.~\ref{pTloss_2760} and \ref{pTloss_5020} are the average transverse energy loss as a function of the vacuum jet $p_T$ in 0-10\% central Pb+Pb collisions (left) at $\sqrt{s}=2.76$ and 5.02 TeV, respectively, and a static medium with a constant temperature $T=0.28$ GeV and finite length (propagation time) $L=4$ fm (right) for three different jet-cone sizes $R=0.3$, 0.4 and 0.5. Without the inclusion of medium response (recoil and ``negative" partons) (black lines with circles) the jet transverse energy loss is significantly larger than that with medium response (red lines with squares).  Inclusion of ``negative" partons increases the jet energy loss only slightly. The inclusion of the medium response (mainly recoil partons) not only reduces the net jet energy loss but also its dependence on the vacuum jet $p_T$, making the $p_T$-dependence much flatter. As we have seen in the last section, this weaker $p_T$-dependence of the jet energy loss is responsible for the  $p_T$-dependence of the jet suppression factor $R_{\rm AA}(p_T)$ given the shape of the vacuum jet spectra in p+p collisions.  The reduction of the jet energy loss due to the inclusion of medium response increases with the jet cone-size, since the energy carried by recoil partons is spread to wide angles away from the jet axis. The radial expansion in the hydrodynamic medium helps to transport recoil partons to a wider angle away from the jet axis. This makes the net jet energy loss more dependent on the jet-cone size as compared to the case of jet propagation in a static medium. This is more so for the effect of ``negative" partons. In all scenarios, the jet energy loss in general decreases with the jet-cone size $R$.

\subsection{Flavor dependence}

It is known that gluons lose more than twice the energy as quarks in a QCD medium and the flavor composition of single inclusive jets in p+p collisions depends on the transverse momentum and colliding energy. The transverse momentum and colliding energy dependence of the average jet energy loss in heavy-ion collisions should also be influenced by the flavor composition of the initial jets. We will examine this in detail here.

\begin{figure}[htbp]
    \includegraphics[width=8.5cm]{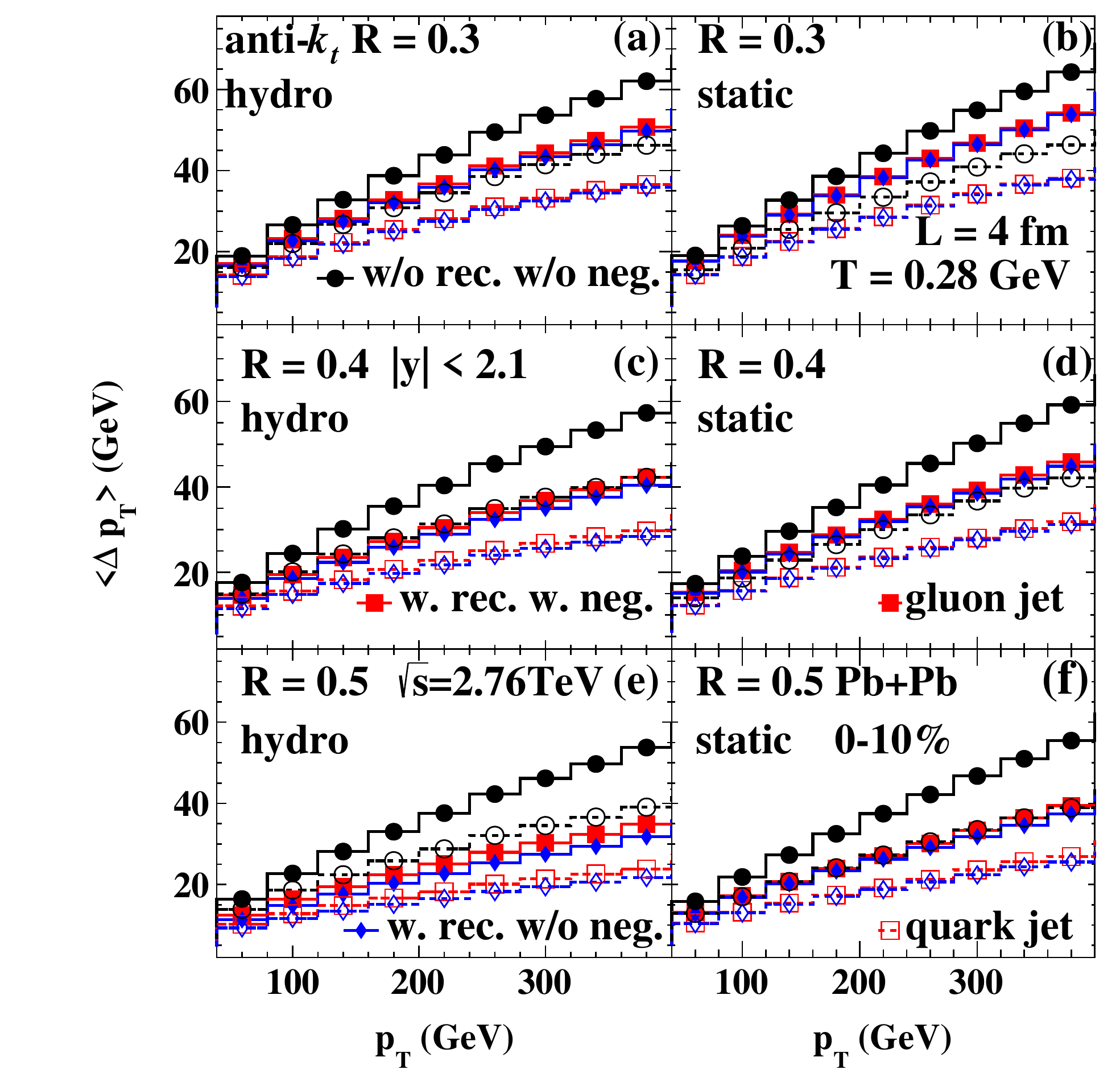}
    \caption{(Color online) The same as Fig.~\ref{pTloss_2760} but for (solid lines) gluon  and (dashed lines) quark jets.}
    \label{pTlossgq_2760}
\end{figure}

\begin{figure}[htbp]
    \includegraphics[width=8.5cm]{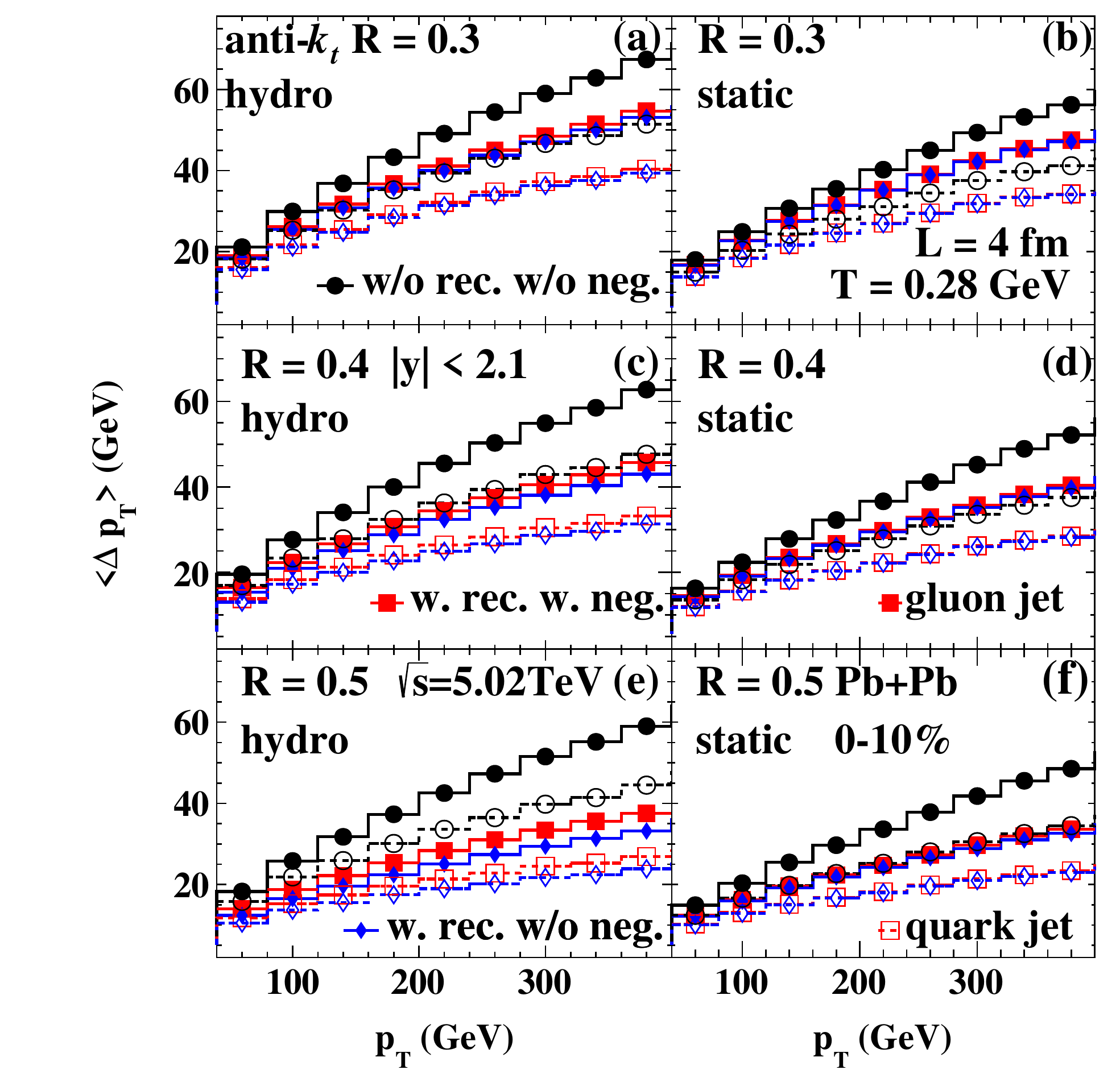}
    \caption{The same as Fig.~\ref{pTloss_2760} except for gluon (solid lines) and quark jets (dashed lines) at $\sqrt{s}=5.02$ TeV.}
    \label{pTlossgq_5020}
\end{figure}

In the high-energy limit when jet shower parton energy is much bigger than the local temperature $E\gg T$, the $t$-channel gluon and quark scattering cross sections can be approximated by their small angle limits,
\begin{eqnarray}
\frac{d\sigma_{ab}}{dq_\perp^2} &\approx& C_{ab} \frac{2\pi\alpha_{\rm s}^2}{(q_\perp^2+\mu_D^2)^2},   \\
&&\left(C_{gg}=\frac{9}{4}, C_{qg}=1, C_{qq}=\frac{4}{9}\right). \nonumber
\label{eq-small-el}
\end{eqnarray}
One can calculate the elastic parton energy loss,
\begin{eqnarray}
\label{eloss}
\frac{dE_{\rm el}^{a}}{dx}&=&\sum_b \int dq_\perp^2 \frac{d^3k}{(2\pi)^3} f_b(k)\frac{q_\perp^2}{2k^0} \frac{d\sigma_{ab}}{dq_\perp^2} \nonumber \\
&\approx&C_a\frac{3\pi}{2} \alpha_{\rm s}^2 T^2 \ln (\frac{s^*}{4\mu _D^2}) ,
\end{eqnarray}
where $s^*\approx 2.6 ET$~\cite{He:2015pra}. Similarly, the jet transport coefficient as defined in Eq.~(\ref{eq-qhat}) is,
\begin{equation}
\label{<qperp>}
\hat q_a
\approx C_a \frac{42 \zeta(3)}{\pi} \alpha_{\rm s}^2T^3 \ln (\frac{s^*}{4\mu _D^2}),
\end{equation}
where $s^*\approx 5.7ET$~\cite{He:2015pra}. Since the radiative gluon spectra in Eq.~(\ref{induced}) is proportional to $\hat q_a$, both the elastic and radiative energy loss of a propagating parton in a QGP medium depend on its color charge, $C_F=4/3$ for a quark and $C_A=3$ for a gluon \cite{He:2015pra,CasalderreySolana:2007sw}.

The net energy loss of a jet in a QGP medium should also depend on the color charge of its originator, though the dependence is weaker than the energy loss of a single parton, since a jet shower contains both quarks (anti-quarks) and gluons whether it is originated from a highly virtual quark or gluon.
In PYTHIA 8 simulations, we tag the flavor of a leading jet in p+p collisions by the flavor of the final parton in the hard $2\rightarrow 2$ processes in the direction of the final jet and assign the same flavor tagging to the final jet after propagation in the QGP medium.  Shown in Figs.~\ref{pTlossgq_2760} and \ref{pTlossgq_5020} are the averaged net jet transverse energy loss as a function of the vacuum jet $p_T$ for gluon (solid lines) and quark jets (dashed lines) with three different jet-cone sizes ($R=0.3$, 0.4, and 0.5) in the static (right) and hydrodynamic QGP medium (left) in 0-10\% Pb+Pb collisions at $\sqrt{s}=2.76$ and 5.02 TeV, respectively. The energy loss of flavor-tagged jets follows the same trend as the flavor-averaged jet energy loss in Figs.~\ref{pTloss_2760} and \ref{pTloss_5020}. Gluon jets however lose more energy than quark jets. The effect of medium response, inclusion of which reduces the net jet energy loss,  is also stronger for gluon jets than quark jets. 

\begin{figure}[htbp]
    \includegraphics[width=8.5cm]{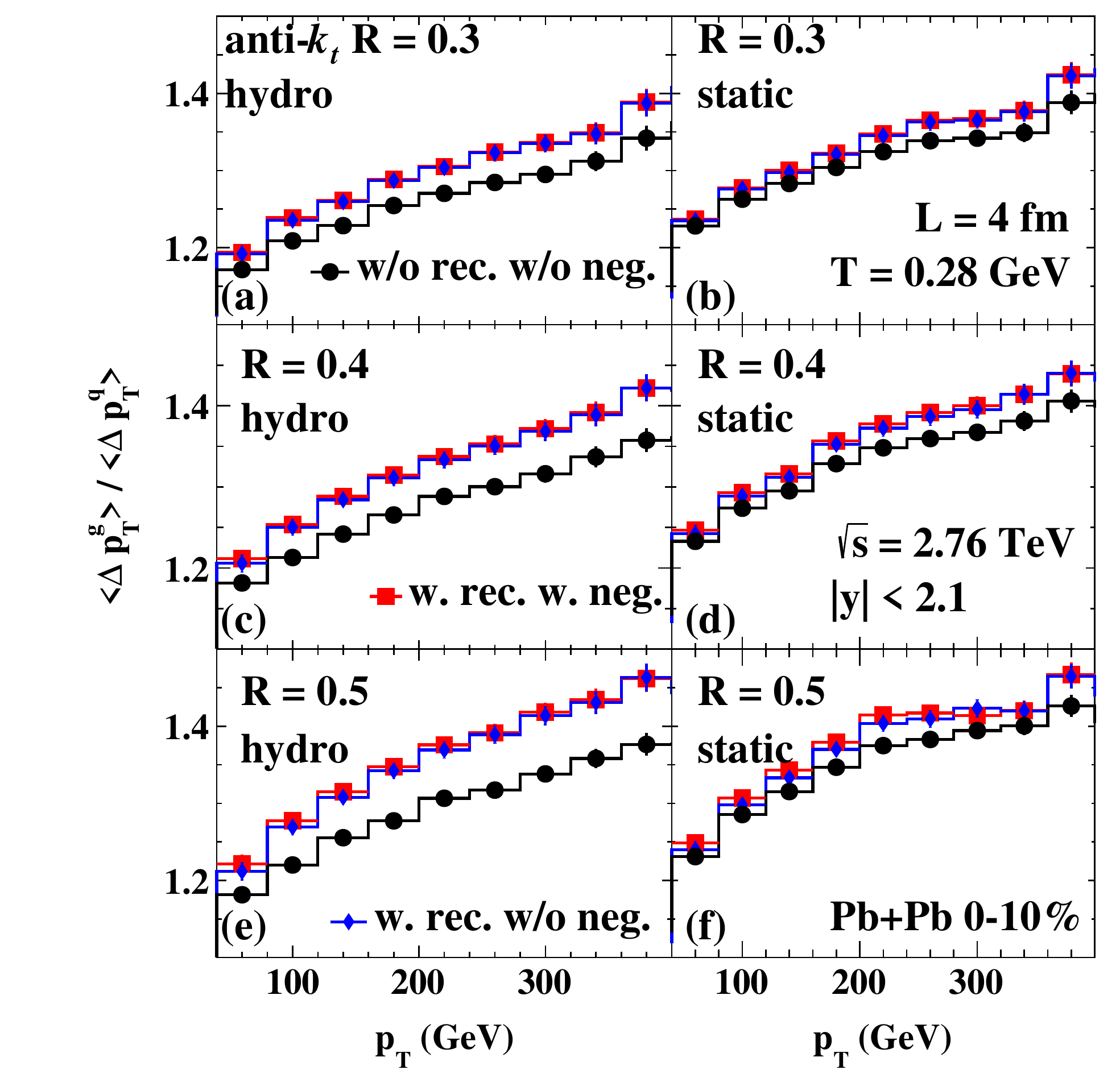}
    \caption{(Color online) LBT results on ratios of energy loss of gluon jets over quark jets in $|y| < 2.1$ as a function of the vacuum jet $p_{T}$ with anti-$k_t$ algorithm and $R = 0.3, 0.4, 0.5$ for [(a), (c), (e)] hydrodynamic background in central 0 - 10\% Pb+Pb collisions at $\sqrt{s} = 2.76$ TeV and [(b), (d), (f)] static medium at $T = 0.28$ GeV with fixed length $L = 4$ fm. Black lines with circles are results without recoil and ``negative" partons, while red lines with squares are with recoil and  ``negative" partons and blue lines with diamonds are with recoil but without ``negative" partons.}
    \label{pTlossgqR_2760}
\end{figure}

\begin{figure}[htbp]
    \includegraphics[width=8.5cm]{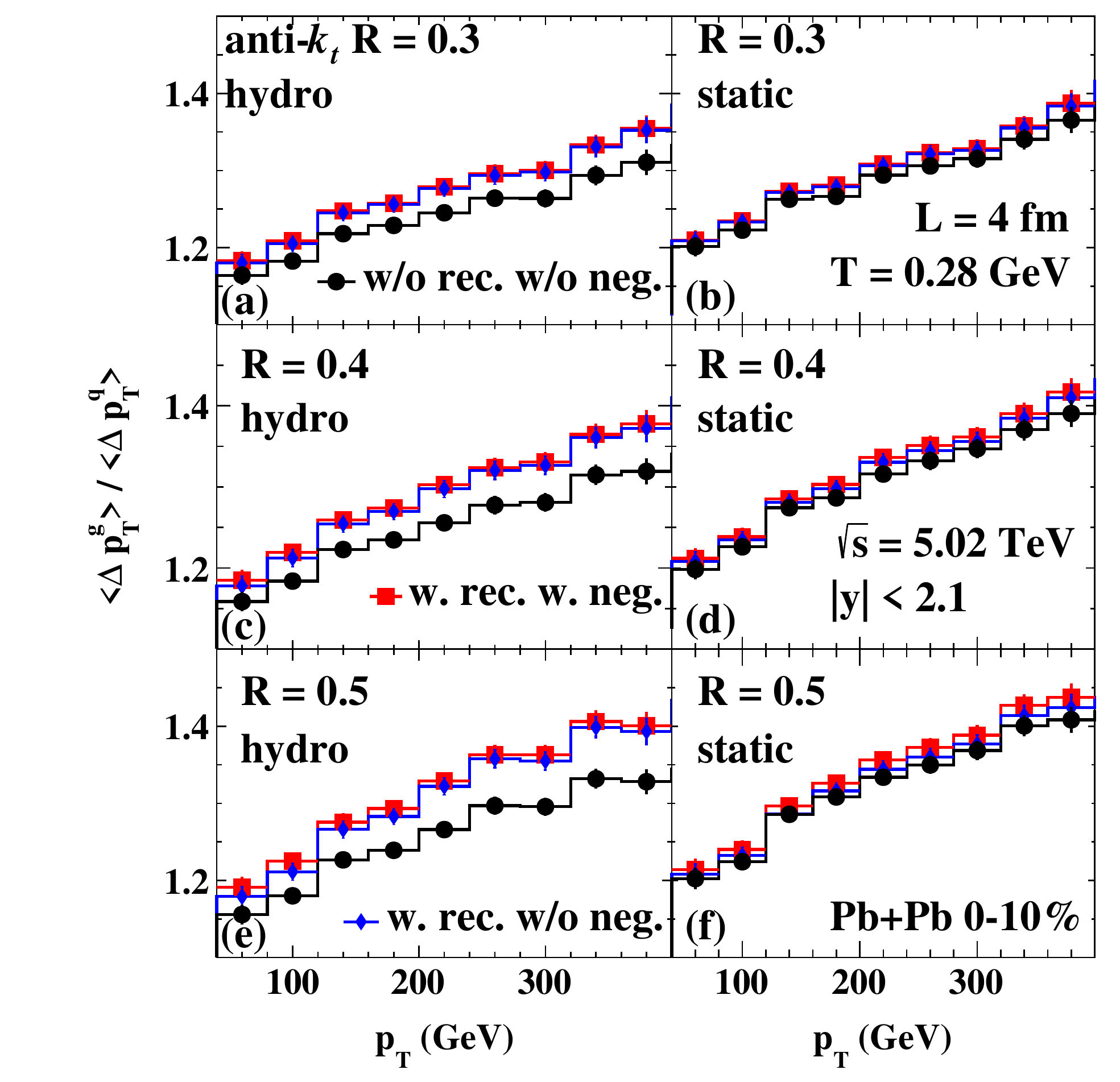}
    \caption{(Color online) The same as Fig.~\ref{pTlossgqR_2760} except for $\sqrt{s}=5.02$ TeV.}
    \label{pTlossgqR_5020}
\end{figure}

To illustrate the difference between gluon and quark jet energy loss, we show in Figs.~\ref{pTlossgqR_2760} and \ref{pTlossgqR_5020} the ratio of gluon to quark jet energy loss  from Figs.~\ref{pTloss_2760} and \ref{pTloss_5020}, respectively. Since jet showers also contain gluons even if they are initiated by a hard quark, the net energy loss of a gluon-tagged jet is always larger than that of a quark-tagged jet but smaller than 9/4, which is the ratio of energy loss of a single gluon and quark, as seen in the LBT calculation. The ratio of gluon and quark-tagged jet energy loss with medium response increases from 1.2 to about 1.4 in the $p_T$ range shown. This means the medium sees more of the jet's original color charge for larger vacuum jet $p_T$.  Without the medium response, the ratio is slightly smaller. This indicates that the effect of medium recoil is bigger for gluon-tagged jets because of their stronger interaction with medium and larger energy loss than quark-tagged jets. The ratio is also slightly influenced by the radial expansion and has moderate dependence on the jet cone-size.

\begin{figure}[!h]
    \includegraphics[width=7.5cm]{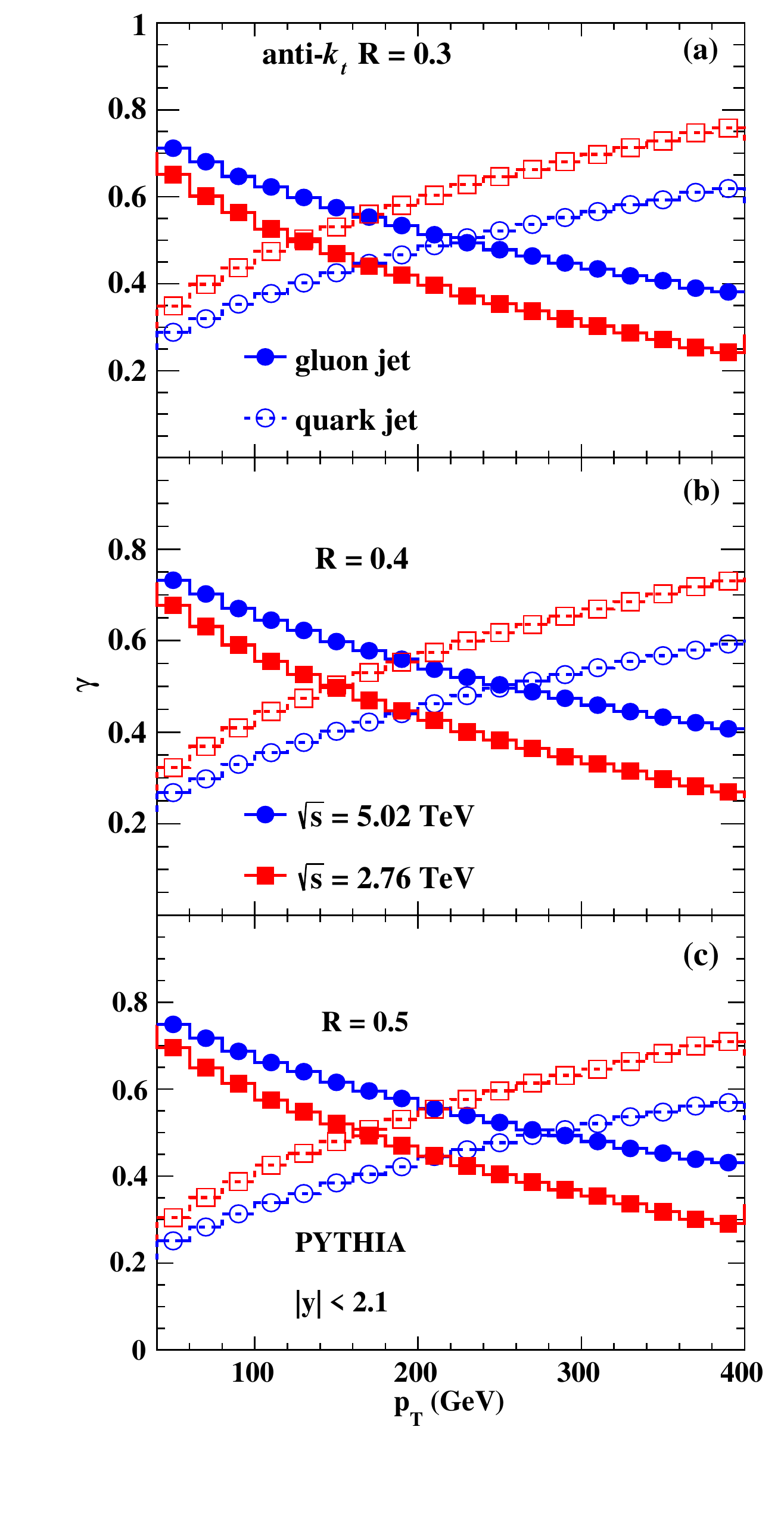}
    \caption{(Color online)  Transverse momentum dependence of the fraction of (solid lines) gluon jet  and (dashed lines) quark jet  within $|y|<2.1$ in p+p collisions at (red squares)  $\sqrt{s} = 2.76$ and  (blue circles) 5.02 TeV  from PYTHIA 8 simulations with anti-$k_t$ and $R = 0.3, 0.4, 0.5$. }
    \label{ngqpT_twoEnergy}
\end{figure}

To better understand the final flavor-averaged jet energy loss, one also needs to know the initial flavor composition of single inclusive jets as reconstructed with FASTJET. Shown in Fig.~\ref{ngqpT_twoEnergy} are fractions of gluon (solid lines) and quark-tagged jets (dashed lines) as a function of the vacuum jet $p_T$ with three different jet-cone sizes ($R=0.3$, 0.4 and 0.5) in p+p collisions at $\sqrt{s}=2.76$ (red squares) and 5.02 TeV (blue circles). The gluon (quark) jet fraction decreases (increases) with the vacuum jet $p_T$ as determined by the parton distributions inside a nucleon. The fractions have almost no dependence on the jet-cone size.  At fixed values of jet $p_T$, the gluon (quark) fraction is bigger (smaller) at higher colliding energy or small parton initial momentum fraction $x_T=p_T/2\sqrt{s}$. We have checked that given these flavor compositions, $\gamma_g(p_T)$ and $\gamma_q(p_T)$ in Fig.~\ref{ngqpT_twoEnergy}, and the flavor-tagged jet energy loss, $\Delta p^g_T(p_T)$ and $\Delta p_T^q(p_T)$ in Figs.~\ref{pTlossgq_2760} and \ref{pTlossgq_5020}, one can recover the inclusive jet energy loss in Figs.~\ref{pTloss_2760} and \ref{pTloss_5020} through
\begin{equation}
 \langle \Delta p_T\rangle =\gamma_g \langle \Delta p^g_T\rangle + \gamma_q \langle \Delta  p_T^q \rangle.
 \end{equation}
 According to this flavor composition, the quark fraction among the inclusive jets increases with $p_T$. Since quark jet energy loss is smaller than gluon jet, the $p_T$-dependence of the effective flavor-averaged jet energy loss for single inclusive jets is weaker than that for flavor-tagged jets (quark or gluon). Together with the effect of recoil partons from medium response, this further weakens the $p_T$-dependence of the effective inclusive jet energy loss and consequently leads to the observed  $p_T$-dependence of the suppression factor $R_{\rm AA}(p_T)$.  As one increases the colliding energy, the gluon jet fraction at fixed $p_T$ increases. This will increase the effective inclusive jet energy loss accordingly. With the increased initial energy density in the bulk medium, the increased inclusive jet energy loss at higher colliding energy is, however, offset by the flatter initial jet spectra and leads to a weak colliding energy dependence of the jet suppression factor. 

\subsection{Rapidity dependence of jet suppression}

\begin{figure}[!h]
    \includegraphics[width=7.5cm]{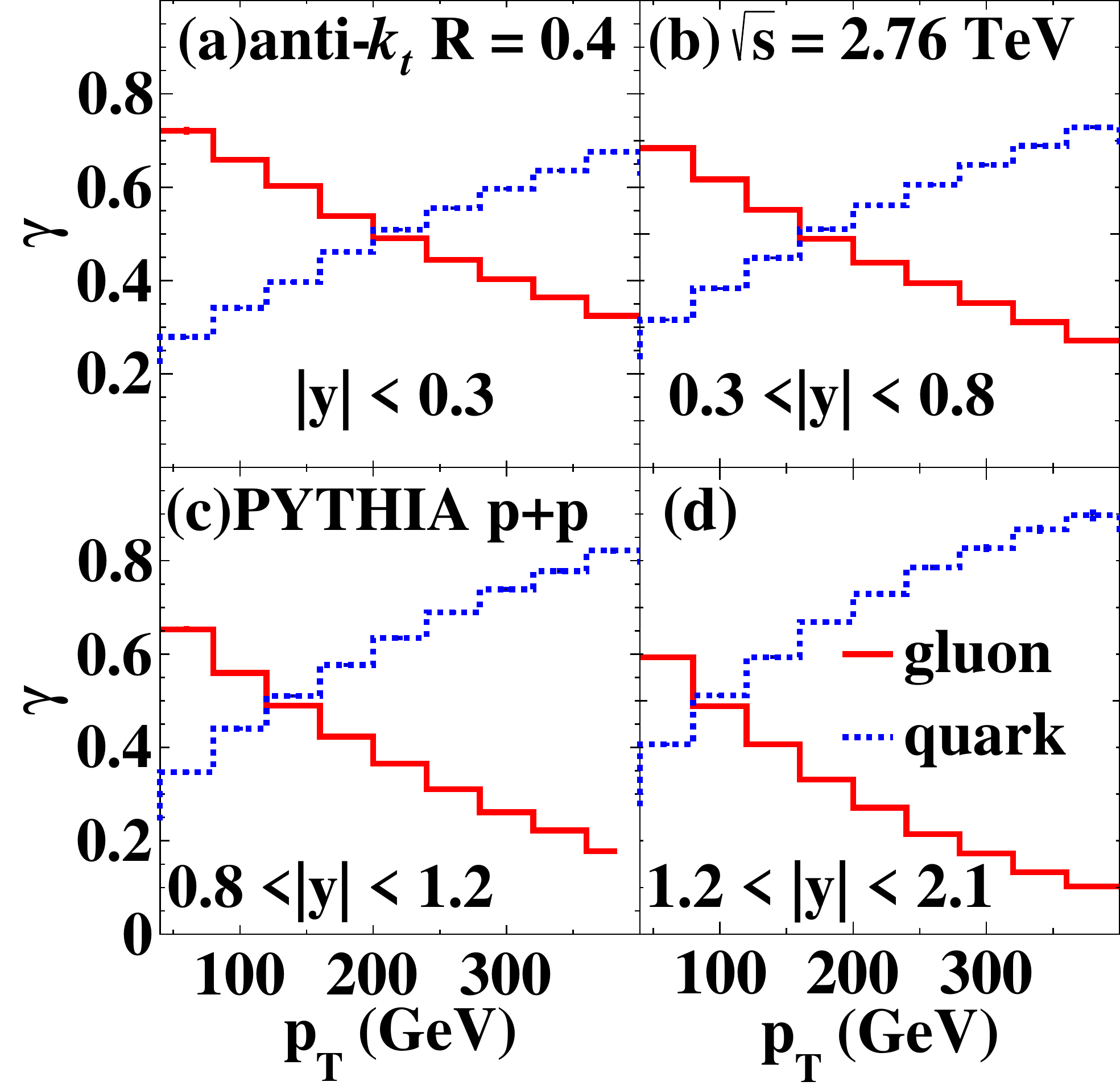}
    \caption{(Color online)  Transverse momentum dependence of the fraction of gluon jet (red solid lines) and quark jet (blue dashed lines) for different jet rapidity $y$ in p+p collisions at $\sqrt{s} = 2.76$ TeV from PYTHIA 8 simulations with anti-$k_t$ and $R = 0.4 $. }
    \label{ngqpT_rapidity}
\end{figure}

\begin{figure}[!h]
    \includegraphics[width=7.5cm]{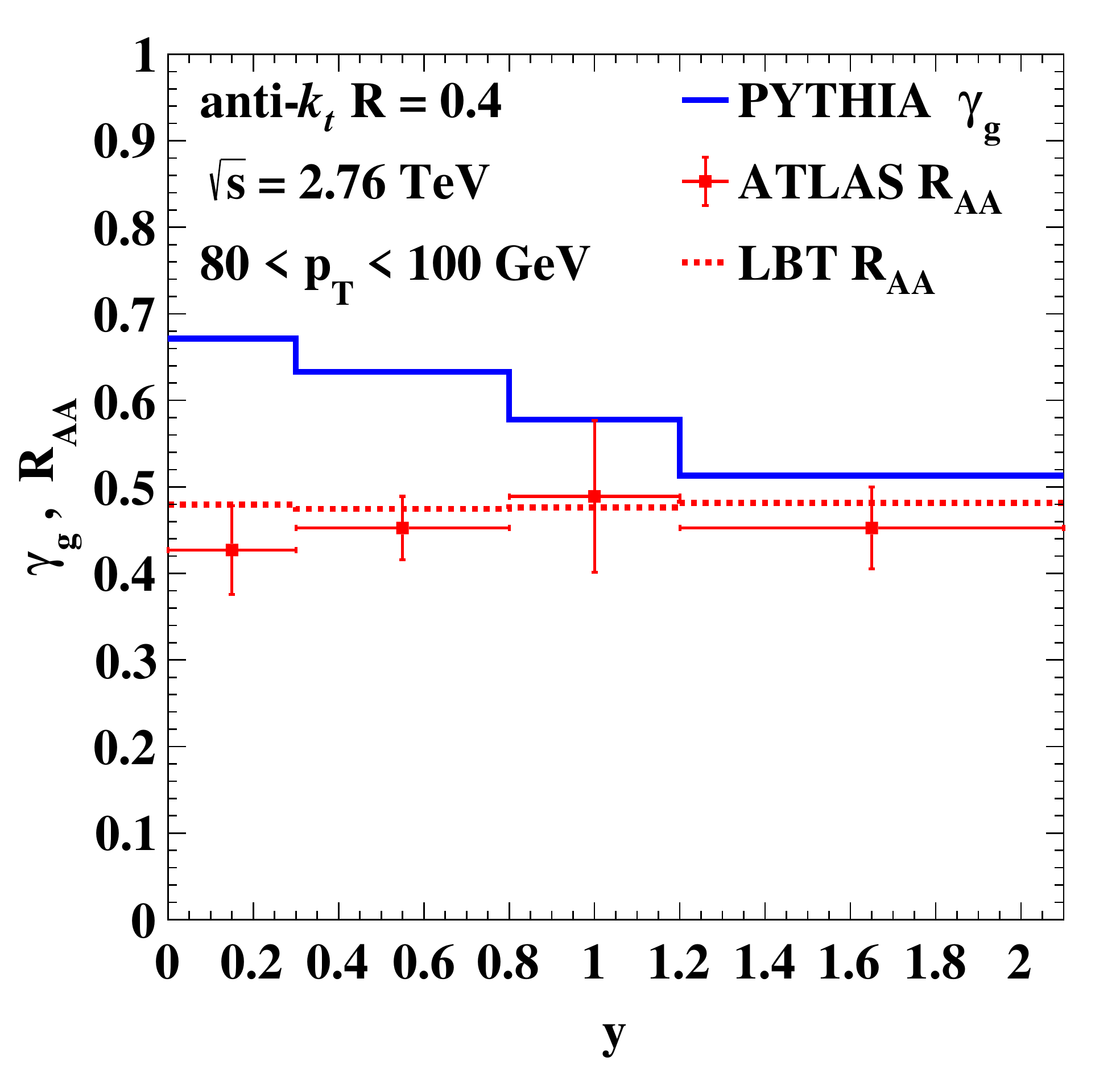}
   \caption{(Color online)  Rapidity dependence of the initial gluon jet fraction $\gamma_g$ (blue solid line)  and jet suppression factor $R_{AA}$ for $80<p_T<100$ (red dashed line) in 0-10\% central Pb+Pb collisions at $\sqrt{s} = 2.76$ TeV from LBT simulations with anti-$k_t$ and $R = 0.4 $. Solid squares are ATLAS data~\cite{Aad:2014bxa} on $R_{AA}$. Note that two different observables, gluon fraction $\gamma_g$ and single jet suppression factor $R_{AA}$, are plotted in this figure.}
    \label{RAA_y}
\end{figure}

The jet flavor composition shown in Fig.~\ref{ngqpT_twoEnergy} are averaged over the central rapidity region $|y|<2.1$ which is determined by the flavor dependence of the parton distribution functions (PDF's) inside a proton and the partonic cross sections.  The flavor dependence, especially gluons versus quarks, of PDF's is known to vary with the momentum fraction $x$ of partons favoring gluons at small  $x$. The jet flavor composition will therefore depend on the rapidity of the final jets.  Shown in Fig.~\ref{ngqpT_rapidity} are the gluon (red solid lines) and quark (blue dashed linee) jet fractions as a function of the vacuum jet $p_T$ with jet-cone size $R=0.4$ in different rapidity bins in p+p collisions at $\sqrt{s}=2.76$ TeV. The gluon (quark) fraction decreases (increases) with rapidity at a fixed value of jet $p_T$. The cross-point where gluon and quark fraction become equal moves to smaller $p_T$ as the rapidity increases.  As an illustration of the rapidity dependence of the flavor composition, we plot in Fig.~\ref{RAA_y} the gluon fraction (blue solid line) as a function of  rapidity for $80<p_T<100$ GeV/$c$  in  p+p collisions at $\sqrt{s}=2.76$ TeV.  It decreases from $\gamma_g=0.68$ at $y=0$ to 0.52 at $y=2.1$. According to Fig.~\ref{pTlossgqR_2760}, gluon jets lose about 1.2 more energy than quark jets for $p_T=80-100$ GeV/$c$. The jet energy loss $\Delta p_T =\Delta p_T^g \gamma_g +(1-\gamma_g)\Delta p_T^q \approx (1+0.2\gamma_g)\Delta p_T^q$ will only decrease by 2.8\% due to the decrease of the gluon fraction from $y=0$ to 2.1. The jet energy loss for both flavors will decrease from central to large rapidity due to the spatial distribution of the bulk medium density. This rapidity dependence of the jet energy loss is offset by the rapidity dependence of the initial jet spectra which become steeper as a function of $p_T$ at large rapidity. The final jet suppression factor $R_{AA}$ will then have a very weak rapidity dependence within the range $0<|y|<2.1$ as shown in Fig.~\ref{RAA_y} for $80<p_T<100$ GeV/$c$ (red dashed line) (see also Fig.~ \ref{RAA_ctrl_rap}) which is consistent with the ATLAS data~\cite{Aad:2014bxa}. Please note that two different observables, gluon fraction $\gamma_g$ and single jet suppression factor $R_{AA}$, are plotted in Fig.~\ref{RAA_y} for convenience.

\subsection{Cone-size dependence of jet suppression}

\begin{figure}[!h]
    \includegraphics[width=7.5cm]{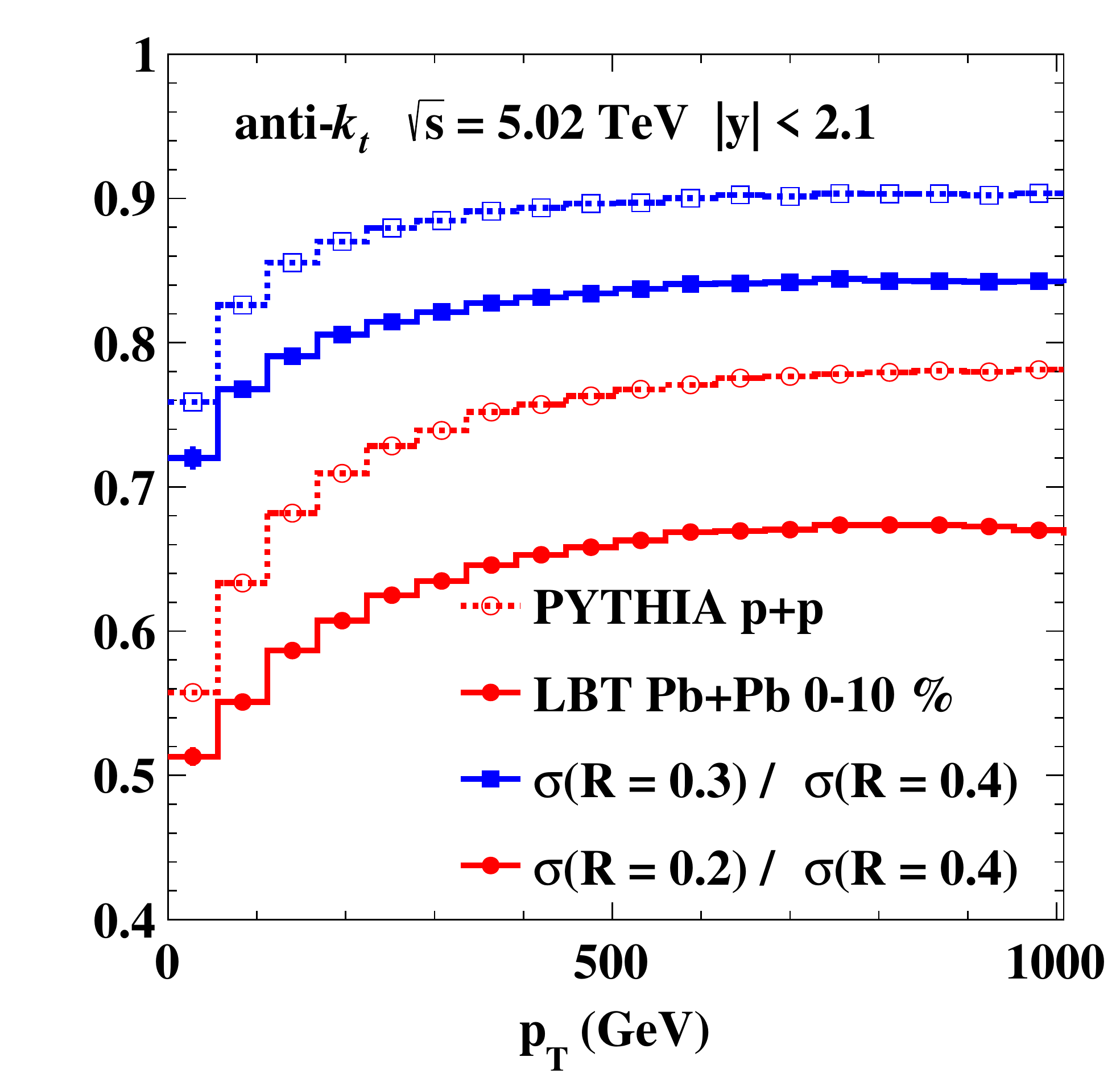}
    \caption{(Color online) Ratios of single inclusive jet spectra with different jet cone-size, $\sigma(R=0.2)/\sigma(R=0.4)$ (lines with squares) and $\sigma(R=0.3)/\sigma(R=0.4)$ (lines with circles), as a function of $p_T$ in (solid) 0-10\% central Pb+Pb and (dashed) p+p collisions at $\sqrt{s}=5.02$ TeV from LBT and PYTHIA 8 simulations, respectively.}
    \label{cone-ratio}
\end{figure}

As we have shown in the above subsections, medium response and radial expansion can both influence the net jet energy loss and lead to a stronger jet-cone size dependence.  The net jet energy loss decreases with the cone size as jets with a bigger cone size will include more medium recoil partons and radiated gluons. This in principle should also lead to a unique cone size dependence of the single inclusive jet suppression, which should also be influenced by the cone size dependence of the single inclusive jet spectra in p+p collisions.  Shown in Fig.~\ref{cone-ratio} are ratios of the single inclusive jet spectra from LBT simulations of 0-10\% central Pb+Pb collisions (solid) as compared to p+p results (dashed) from  PYTHIA 8 with different cone sizes in the central rapidity region at $\sqrt{s}=5.02$ TeV.  One can see that single inclusive jet spectra are in general smaller for smaller jet-cone size. The bigger energy loss for jets with smaller jet-cone size will further reduce the spectra relative to that with a bigger jet-cone size. Though the magnitude of the jet spectra decreases with smaller jet-cone size, the shape of the spectra is actually flatter [$\sigma(R=0.2)/\sigma(R=0.4)$ and $\sigma(R=0.3)/\sigma(R=0.4)$ both increase with $p_T$].  Since the net jet energy loss increases with smaller jet-cone size, the corresponding jet suppression should be stronger (smaller values of $R_{\rm AA}$), which in turn should be off-set somewhat by the flatter jet spectra in vacuum. 

\begin{figure}[!h]
    \includegraphics[width=7.5cm]{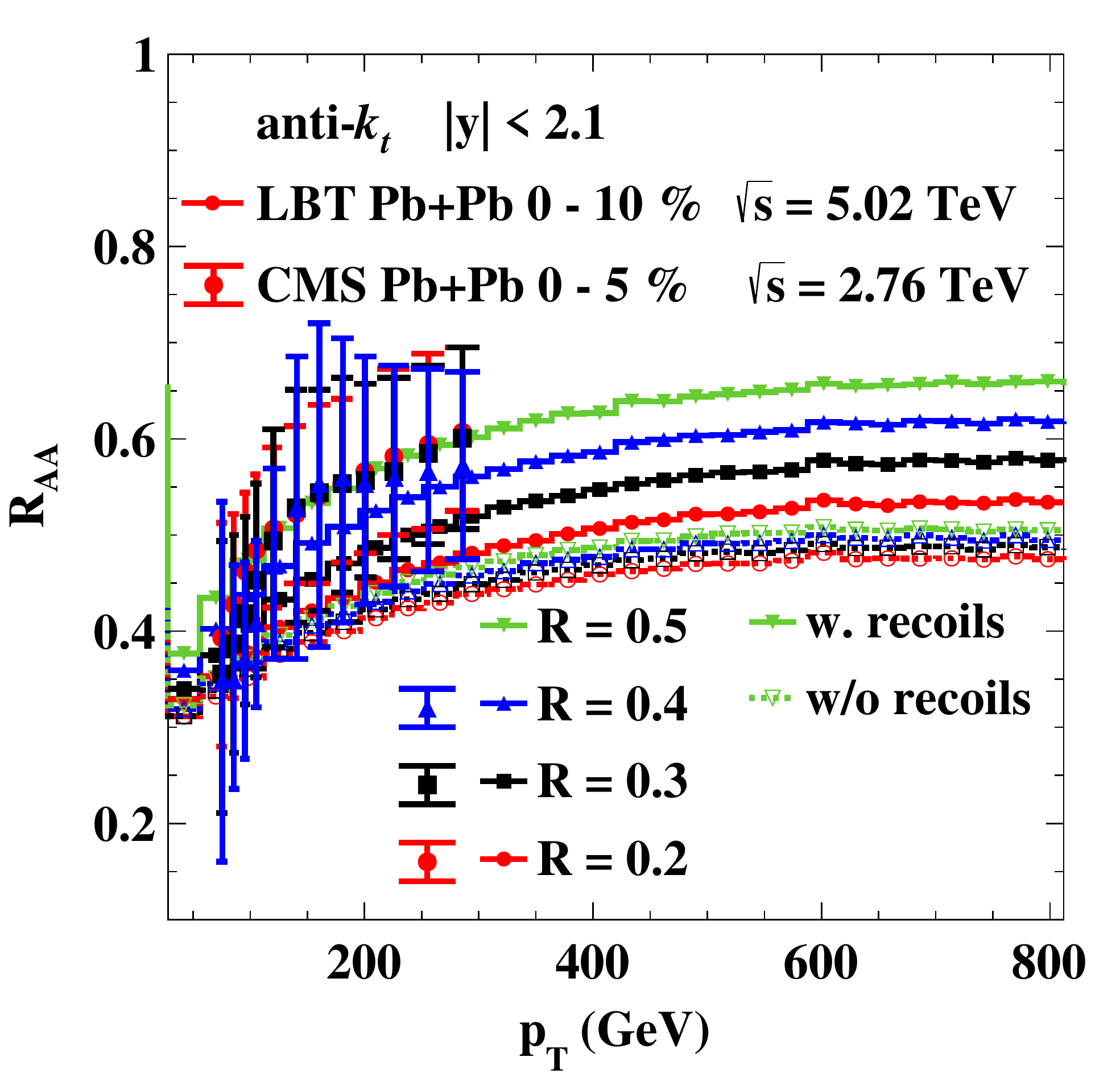}
    \caption{(Color online) Suppression factor of single inclusive jet spectra $R_{\rm AA}$ as a function of $p_T$ in central rapidity region of 0-10\% Pb+Pb collisions at $\sqrt{s}=5.02$ TeV from LBT with (solid) and without medium recoil (including ``negative" partons) (dashed) for different jet-cone sizes, $R$=0.5, 0.4, 0.3 and 0.2 as compared to CMS data \cite{Khachatryan:2016jfl} in 0-5\% Pb+Pb collisions at $\sqrt{s}=2.76$ TeV.}
    \label{RAA-cone}
\end{figure}

Shown in Fig.~\ref{RAA-cone} are LBT results on the single jet suppression factor with (solid) and without medium recoil (including ``negative" partons) (dashed) as a function of $p_T$ in central rapidity region of 0-10\% Pb+Pb collisions at $\sqrt{s}=5.02$ TeV for different jet-cone sizes, $R$=0.5, 0.4, 0.3 and 0.2. We observe that the suppression factor increases with the jet-cone size as the net jet energy loss gets smaller for bigger jet-cone size.  Without medium recoil, the suppression factors are not only significantly smaller due to increased energy loss but also much less sensitive to the jet-cone size.  The jet suppression as measured by CMS experiment~\cite{Khachatryan:2016jfl} for Pb+Pb collisions at $\sqrt{s}=2.76$ TeV show almost no jet-cone size dependence. However, the systematic uncertainties are too big to see the predicted jet-cone size dependence from LBT simulations shown. Similar behavior was also predicted in Refs.~\cite{Vitev:2008rz,Vitev:2009rd,Zapp:2012ak,Kang:2017frl}. But the $p_T$-dependence is different in LBT because of the influence of medium response and radial expansion. More precision measurements of the cone size dependence of the jet suppression can therefore elucidate the underlying processes responsible for the final jet suppression.

\section{Predictions at RHIC}
\label{rhicpredict}
\begin{figure}[htbp]
    \centering
    \includegraphics[width=7.5cm]{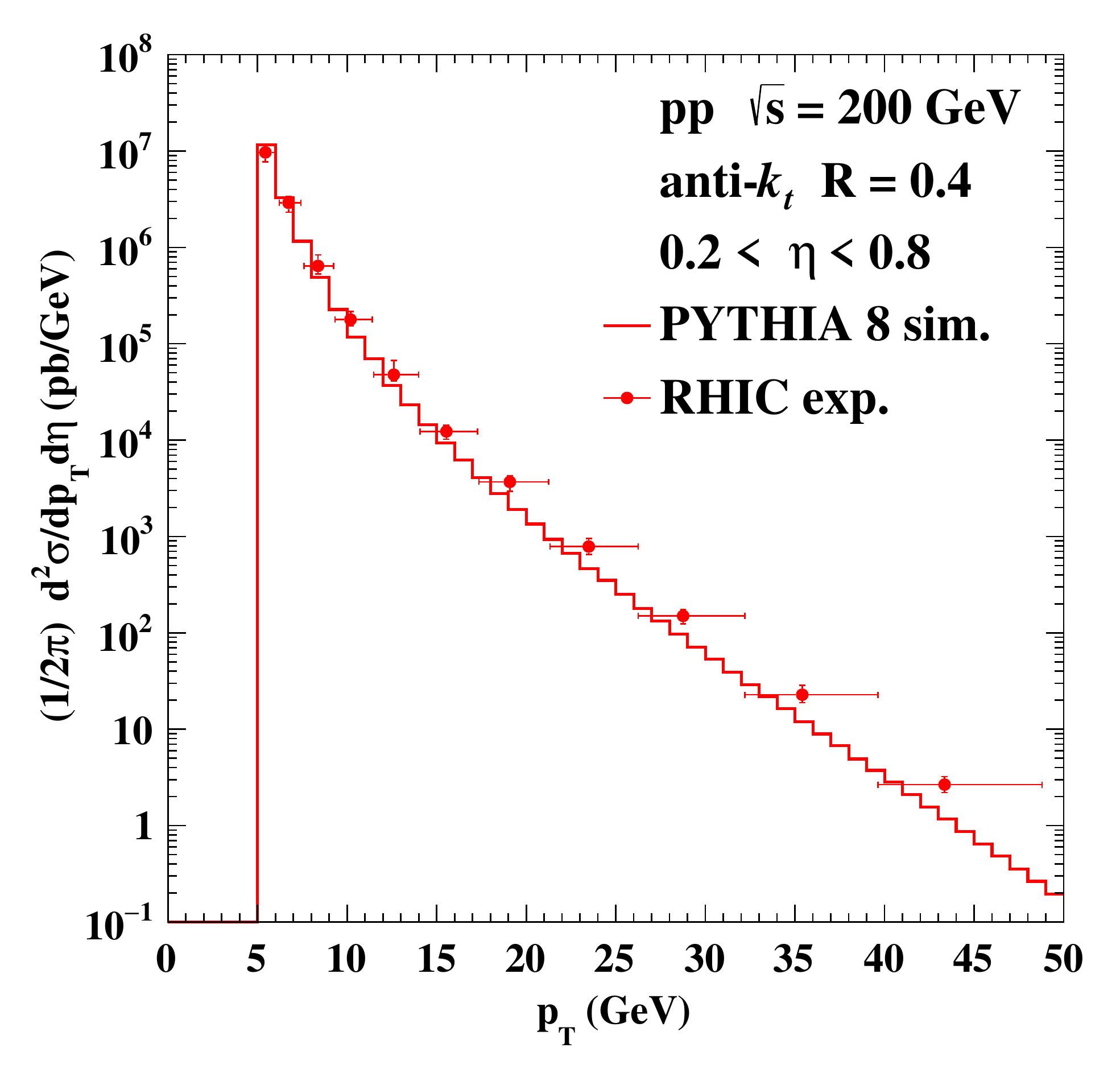}
    \caption{(Color online) PYTHIA 8 result of the inclusive jet differential cross section as a function of $p_{T}$ in the central rapidity of p+p collisions at $\sqrt{s} = 200$ GeV with anti-$k_{t}$ algorithm and jet-cone radius R = $0.4$ as compared to STAR data \cite{Abelev:2006uq}.}
    \label{jetCS200}
\end{figure}

\begin{figure}[htbp]
    \centering
    \includegraphics[width=7.5cm]{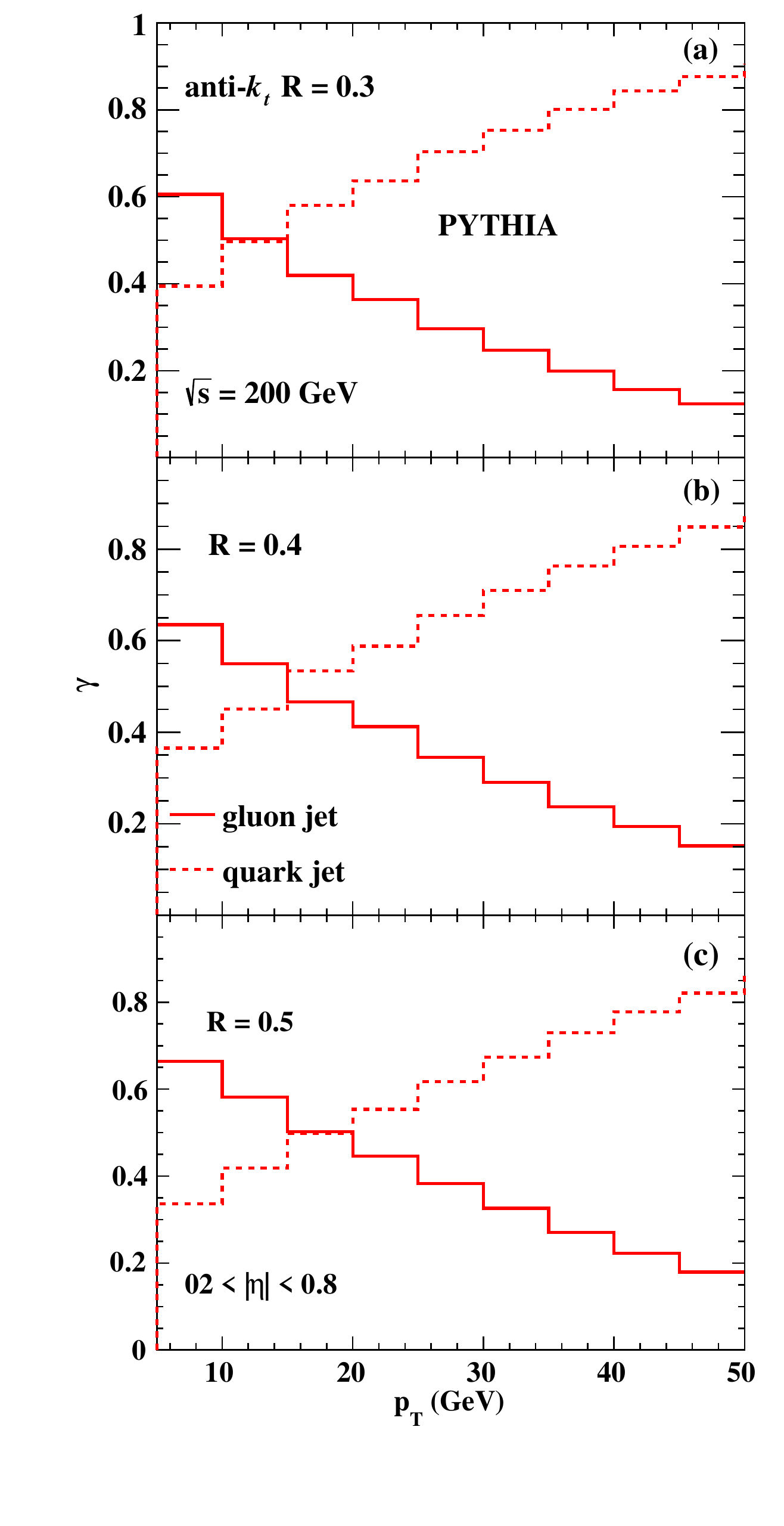}
    \caption{(Color online) Transverse momentum dependence of the number fraction of (solid) gluon  and (dashed) quark jets  in p+p collisions at $\sqrt{s} = 200$ GeV from PYTHIA 8 with anti-$k_t$ and jet-cone sizes $R = 0.3, 0.4$ and 0.5.}
    \label{ngqpT_200}
\end{figure}

\begin{figure}[htbp]
    \includegraphics[width=7.5cm]{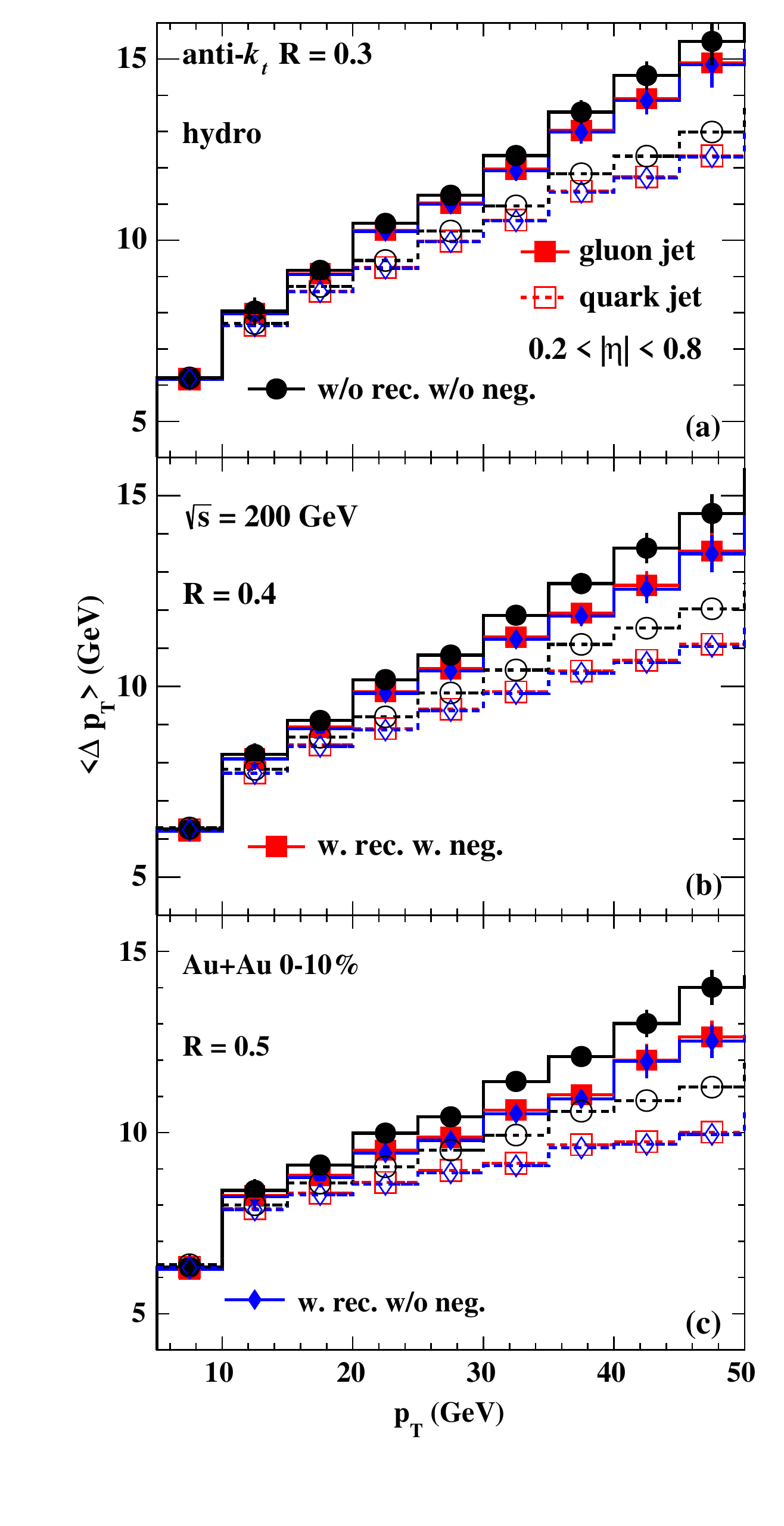}
    \caption{(Color online) LBT results on the average jet transverse energy loss of (solid) gluon  and (dashed)  quark jets within $|y| < 2.1$ with anti-$k_t$ algorithm and jet-cone sizes $R = 0.3, 0.4, 0.5$ as a function of the vacuum jet $p_{T}$ in $0 - 10 \%$ central Au+Au collisions at $\sqrt{s} = 200$ GeV. Black lines with circles  are without recoil and ``negative" partons, while red lines with squares are with recoil and ``negative" partons and blue lines with diamonds are with recoil but without ``negative" partons.}
    \label{pTlossgq_200}
\end{figure}

\begin{figure}[htbp]
    \includegraphics[width=7.5cm]{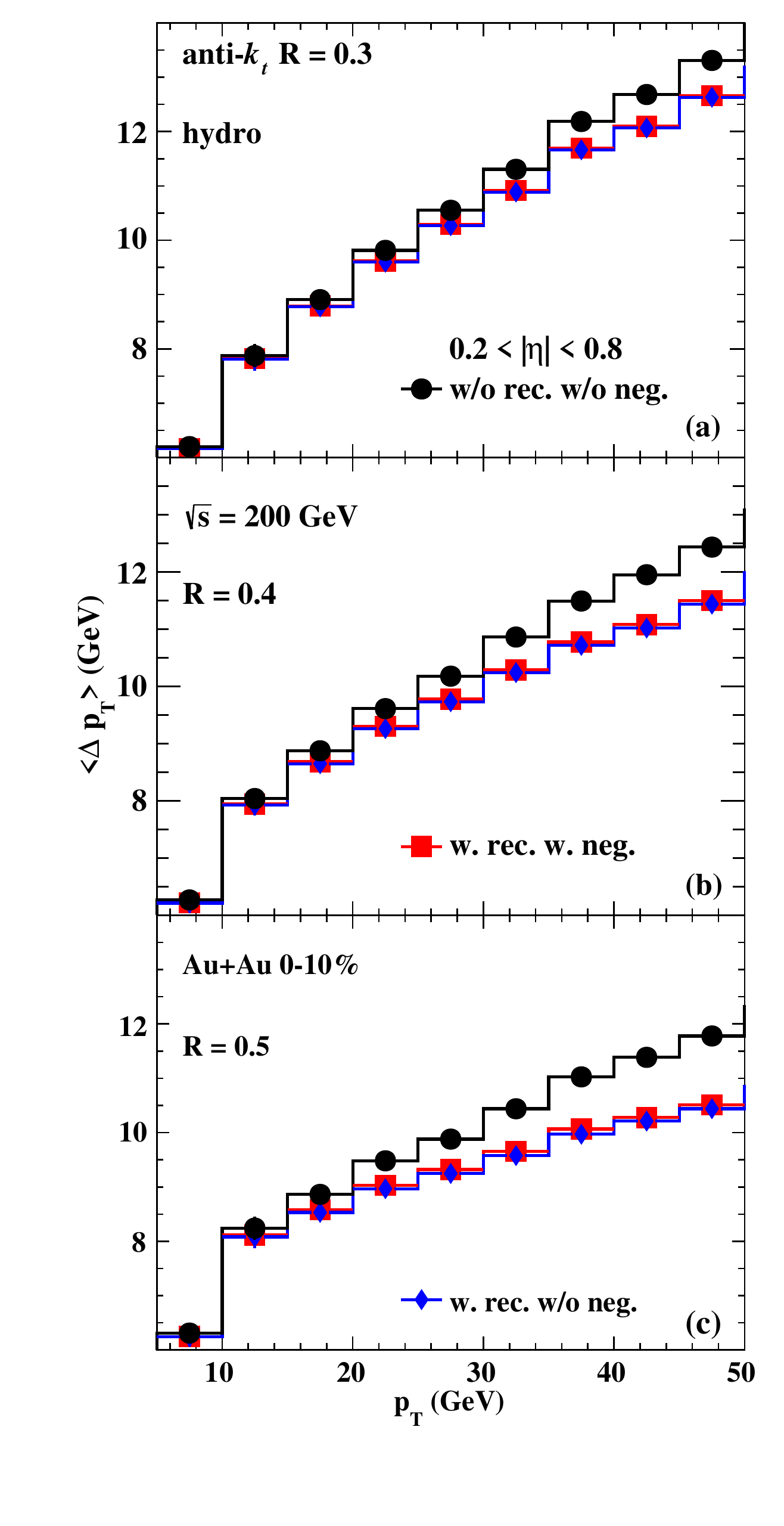}
    \caption{(Color online) The same as Fig.~\ref{pTlossgq_200} except for flavor-averaged jet transverse energy loss.    }
    \label{pTloss_200}
\end{figure}

\begin{figure}[htbp]
        \centering
        \includegraphics[width=8.5cm]{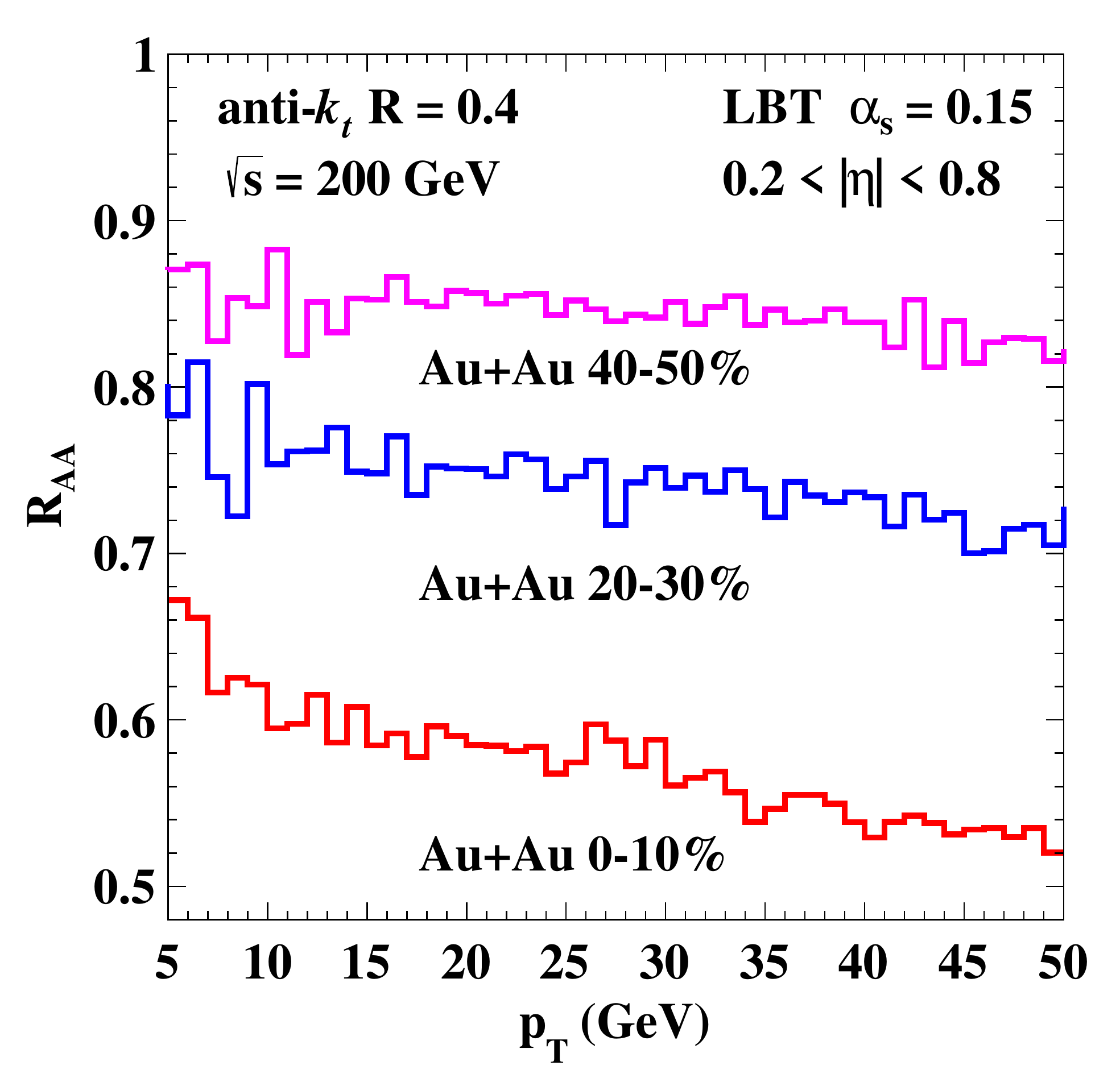}
       \caption{(Color online) LBT predictions for $R_{\rm AA}$ of single inclusive jet spectra in Au+Au collisions at $\sqrt{s}=200$ GeV with three different centralities.}
           \label{RAA_200}
\end{figure}

As we have shown in this study, the transverse momentum dependence of the single inclusive jet suppression factor in heavy-ion collisions is determined mainly by the $p_T$-dependence of the jet energy loss and the shape of the initial single inclusive jet spectra in p+p collisions. Since the single inclusive jet spectra at RHIC energy $\sqrt{s}=200$ GeV is much steeper in the $p_T$ range available as shown by PYTHIA 8 results and STAR experimental data~\cite{Abelev:2006uq} in Fig.~\ref{jetCS200}, the single inclusive jet suppression factor at RHIC should have different transverse momentum dependence from that at LHC, depending on the $p_T$-dependence of the jet energy loss. While fractions of quark and gluon-initiated jets are about the same at around $p_T$= 20 GeV/$c$, jets become mostly quark-dominated at large $p_T$ at RHIC as shown by PYTHIA 8 results in Fig.~\ref{ngqpT_200}. The net energy loss for quark and gluon-initiated jets in the RHIC $p_T$ range is however very similar as shown in Fig.~\ref{pTlossgq_200}. The effect of jet-induced medium response is also much smaller in this $p_T$ range and the jet energy loss has a weak dependence on jet cone size, both due to a shorter duration of the QGP phase in central Au+Au collisions at RHIC. The net jet energy loss as shown in Fig.~\ref{pTloss_200} has a weaker transverse momentum and jet-cone size  dependence as compared to that at LHC. The combined effect of the steep initial jet spectra at RHIC and weak transverse momentum dependence of the jet energy loss in the $p_T$ range leads to the single inclusive jet suppression factor that actually decreases slightly with the final jet transverse momentum as shown in Fig.~\ref{RAA_200} for Au+Au collisions with three different centralities at $\sqrt{s}=200$ GeV.  This is quite different from the $p_T$-dependence of the jet suppression factor at the LHC  that increases with $p_T$, though weakly.  This unique colliding energy and transverse momentum dependence of the single inclusive jet suppression at RHIC will be important to verify and one can directly infer the $p_T$-dependence of jet energy loss given the measured initial jet production spectra in p+p collisions at the same energy~\cite{He:2018gks}.

\section{Conclusions}
\label{summary}

We have carried out a systematic study of jet energy loss and single inclusive jet suppression in high-energy heavy-ion collisions within the LBT model with CLVisc (3+1)D event-by-event hydrodynamic evolution of the bulk medium which is constrained by the bulk hadrons spectra. The LBT model can describe well the dependence of the jet suppression factor $R_{\rm AA}(p_T)$ on the colliding energy, centrality, transverse momentum and rapidity as measured by experiments at LHC. While the average net jet energy loss with a given jet-cone size in Pb+Pb collisions at $\sqrt{s}=5.02$ TeV is larger than that at $\sqrt{s}=2.76$ TeV due to the increased initial bulk medium density and larger fraction of gluon-initiated jets, the final jet suppression factor $R_{\rm AA}(p_T)$ at $\sqrt{s}=5.02$ TeV is actually comparable or even slightly larger than that at $\sqrt{s}=2.76$ TeV. This colliding energy dependence is mainly determined by the initial jet production spectra in p+p collisions which are harder at 5.02 TeV as compared to that at 2.76 TeV. The weak transverse momentum dependence of jet suppression factor at both energies is dictated by the initial jet production spectra, $p_T$-dependence of the net jet energy loss and the jet energy loss fluctuations. We have analyzed the net jet energy loss and its $p_T$-dependence within the LBT model in detail. We found that it is influenced by inclusion of jet-induced medium response, radial expansion and jet flavor (quark and gluon) composition, all leading to a weaker $p_T$-dependence of the averaged jet energy loss. The inclusion of jet medium response and influence of radial expansion also lead to a stronger cone-size dependence of the net jet energy loss. We have shown that this will also lead to a unique cone-size dependence of the single jet suppression.

We have also provided predictions for the single inclusive jet suppression factor in Au+Au collisions at the RHIC energy $\sqrt{s}=200$ GeV. Because of the  steeper initial jet production spectra, we predict that the jet suppression factor at RHIC actually decreases slightly with $p_T$ in the $p_T<50$ GeV/$c$ range,
though the $p_T$-dependence of net jet energy loss is weaker than that at LHC. Such unique energy and $p_T$-dependence of the jet suppression factor is a direct consequence of the $p_T$-dependence of jet energy loss given the measured initial jet production spectra in p+p collisions. Extraction of the $p_T$-dependence of jet energy loss and energy loss fluctuations will provide an important link between experimental measurement of jet suppression and jet transport properties in quark-gluon plasma in high-energy heavy-ion collisions.

\clearpage

\begin{acknowledgments}

This work was supported in part by the National Science Foundation of China  under Grant No. 11221504, by the Major State Basic Research Development Program in China under Grant No. 2014CB845404, by the Director, Office of Energy Research, Office of High Energy and Nuclear Physics, Division of Nuclear Physics, of the U.S. Department of Energy under Contract No. DE-AC02-05CH11231 and No. DE-SC0013460, and by the US National Science Foundation within the framework of the JETSCAPE collaboration, under Grant No. ACI-1550228 and No. ACI-1550300. This research used GPU workstations at CCNU and computer resources of the National Energy Research Scientific Computing Center (NERSC), a U.S. Department of Energy Office of Science User Facility operated under Contract No. DE-AC02-05CH11231.

\end{acknowledgments}

\bibliography{xnwrefs}

\end{document}